\documentclass[lettersize,journal]{IEEEtran}
\usepackage{amsmath,amsfonts}
\usepackage{algorithmic}
\usepackage{algorithm}
\usepackage{array}
\usepackage[caption=false,font=normalsize,labelfont=sf,textfont=sf]{subfig}
\usepackage{textcomp}
\usepackage{stfloats}
\usepackage{url}
\usepackage{verbatim}
\usepackage{graphicx}   
\usepackage{cite}
\usepackage{amssymb}
\usepackage{color}
\usepackage{amsthm}
\usepackage{tabularx}
\usepackage{mathrsfs}
\usepackage{multirow}
\usepackage{makecell}
\usepackage{multicol}
\usepackage{stfloats}
\usepackage{float}
\usepackage{color}
\usepackage{booktabs}
\usepackage{colortbl}
\usepackage{bbding}
\usepackage{multirow}
\begin{document}

\title{Distributionally Robust Day-ahead Scheduling for Power-traffic Network under a Potential Game Framework}

\author{Haoran Deng, Bo Yang, Chao Ning, Cailian Chen and Xinping Guan
\thanks{\emph{Corresponding author: Bo Yang.}}
\thanks{The authors are with the Department of Automation, Shanghai Jiao Tong University, Shanghai 200240, China, the Key Laboratory of System Control and Information Processing, Ministry of Education of China, Shanghai 200240, China, and also with Shanghai Engineering Research Center of Intelligent Control and Management, Shanghai 200240, China.}}

\markboth{Journal of \LaTeX\ Class Files,~Vol.~XX, No.~XX, XX~XXXX}%
{Shell \MakeLowercase{\textit{et al.}}: A Sample Article Using IEEEtran.cls for IEEE Journals}


\maketitle

\begin{abstract}
Widespread utilization of electric vehicles (EVs) incurs more uncertainties and impacts on the scheduling of the power-transportation coupled network. This paper investigates optimal power scheduling for a power-transportation coupled network in the day-ahead energy market considering multiple uncertainties related to photovoltaic (PV) generation and the traffic demand of vehicles. The crux of this problem is to model the coupling relation between the two networks in the day-ahead scheduling stage and consider the intra-day spatial uncertainties of the source and load. Meanwhile, the flexible load with a certain adjustment margin is introduced to ensure the balance of supply and demand of power nodes and consume the renewable energy better. Furthermore, we show the interactions between the power system and EV users from a potential game-theoretic perspective, where the uncertainties are characterized by an ambiguity set. In order to ensure the individual optimality of the two networks in a unified framework in day-ahead power scheduling, a two-stage distributionally robust centralized optimization model is established to carry out the equilibrium of power-transportation coupled network. On this basis, a combination of the duality theory and the Benders decomposition is developed to solve the distributionally robust optimization (DRO) model. Simulations demonstrate that the proposed approach can obtain individual optimal and less conservative strategies.
\end{abstract}


\section{Introduction}
\IEEEPARstart{I}{n} the past decade, transportation electrification grows rapidly as a result of the increasing electric vehicles (EVs) charging demand with a positive role in alleviating environmental pollution of EVs for metropolis \cite{J. A. P. Lopes_Integration}. Public fast charging stations (FCSs) become the main source of the energy for EVs gradually. The spatial-temporal distribution characteristics of EVs flow can lead to the uncertainty of charging load in the power network (PN), which will result in a direct impact on the power scheduling strategy \cite{S. Xie_Two-stage}.  In addition, with the large-scale application of renewable energy, the stability of the power network is impacted seriously by the uncertainty of generated energy.  As a result, the day-ahead energy dispatch of the power-transportation coupled network gains more and more research interest. Under this context, due to the complex coupling characteristics of the two networks, an urgent need for day-ahead energy scheduling while ensuring individual optimality of this coupled network considering multiple uncertainties is required.

There are the characteristics of mutual coupling between the transportation network (TN) and the power distribution network. Specifically, the traveling and charging plans of EVs can be affected by the congestion time and FCS electricity price strategies, and in turn, the EV route choices and traffic flow distribution can redistribute the charging load of PN. In the power-transportation coupled network engineering literature \cite{F. He_Optimal, X. Dong_A charging, K. Mahmud_A review}, the system modeling considering the interaction of coupled network in an equilibrium modeling framework has been widely researched. In addition, the traffic side is impacted by electricity price on traffic assignment, and further, the power flow is redispatched according to the temporal and spatial fluctuations of the charging load caused by the price. As a result, the two networks achieve a dynamic balance of supply and demand through the electricity price decisions. In this context, researchers proposed some schemes towards a socially optimum operating point taking into account travel time and the variations of electricity prices for battery charging in spatial space, while the traffic congestion is considered simultaneously \cite{M. Alizadeh_Optimal,W. Wei_Network,W. Gan_Coordinated}. In addition, EV drivers in the coupled network are strategic and not just electricity price takers. In this connection, game theory provides an effective method to investigate the coupled relation between traffic assignment and power dispatch. Extensive game-theoretic approaches about the traffic assignment \cite{J.R. Correa_Wardrop}, \cite{D. Acemoglu_Competition} and EV charging problems \cite{W. Lee_Electric,C. Wu_Vehicle,W. Tushar_Economics} have been developed. Among the game theoretic approach, there is a special one called potential game, which can reflect incentive of all individuals to change their strategy by using a single global function called the potential function. Thus the individual objectives can be achieved by optimizing the potential function. The authors in \cite{S. Bahrami_A Potential} and \cite{Z. Zhou_Power-traffic} analyzed the interdependency between the EV travelers and power operators by establishing the potential game-theoretic model to achieve the co-optimization of the payoff for the two networks. Nevertheless, the electricity is not generated instantaneously and random demands of FCSs must be met immediatly. Therefore, advance planning is required to ensure the intra-day balance of supply and demand. In the day-ahead power scheduling problem, the spatial-temporal interaction characteristics and the individual selfishness of the coupled network need to be considered from the game perspective. However, the intra-day uncertainties exist in the coupled network, which need to be taken into account in the day-ahead scheduling. Authors in \cite{M. S. Javadi_A two-stage} and \cite{A. Rezaee Jordehi_Two-stage} considered the uncertainties of demands and renewable power in the day-ahead scheduling problem of PN to reduce intra-day scheduling costs effectively.

For the power-transportation coupled network, uncertainty is generally considered in a stochastic optimization (SO) or robust optimization (RO) framework, which is computationally tractable. In \cite{A. Rezaee Jordehi_Two-stage,S. A. Mansouri_A multi-stage,S. A. Mansouri_A sustainable}, the co-operation problem of MGs was formulated as a two-stage SO problem which was reformulated as a mixed-integer linear programing (MILP) framework. Authors in \cite{W. Wei_Robust} and \cite{C. He_Co-optimization} established a two-stage RO model considering the uncertainties of wind power generation and electric demand for the coupled network, which was solved by a delayed constraint generation algorithm and the column-and-constraint generation (C\&CG) method, respectively. In \cite{S. Dehghan_Robust}, the authors introduced an adaptive three-stage robust model considering the uncertain load and power production with bounded intervals. However, the accurate probability distribution of uncertain parameters in SO is unobtainable in practical applications. Meanwhile, the RO model in the above literature considering the upper/lower boundary values of uncertain variables can result in over-conservative decision-making \cite{R. Zhou_Distributional}, due to the probability distribution information of uncertain variables is not incorporated in the optimization problem.

In the actual system decision-making problem, it is difficult to obtain the probability distribution of uncertain parameters accurately, and a part of the probability distribution information can be inferred by decision makers from limited data. In this regard, researchers have proposed a distributionally robust optimization (DRO) method to analyze the uncertainties in system optimization problems, and the risk aversion caused by the unknown probability distribution of uncertainty can be characterized more adequately. The construction of ambiguity set plays a crucial role in the DRO problem. In general, ambiguity sets are divided into the moment information \cite{A. Zare_A distributionally,D. Pozo_An ambiguity-averse,Y. Zhang_Distributionally}, statistical distance \cite{Y. Cao_Optimal,C. Wang _ Risk-based}, and data-driven approaches \cite{C. Zhao_Data-driven,C. Ning _Data-driven}. Among them, the moment-based ambiguity set is commonly applied, since it possesses better solvability and can reflect the characteristics of the uncertainty probability distribution accurately.

Based on the existing literature, the key of modeling is to construct the coupling relation of the two networks considering source-load uncertainties in this study. In the optimization problem modeling, the decision-making selfishness of each individual in the coupled network should be considered to ensure that the individual revenue is optimal. Meanwhile, in the day-ahead scheduling optimization problem, it is necessary to ensure that the obtained strategies can provide appropriate margin for the intra-day selfish behavior of individuals. Therefore, the application of the game theory is essential. In addition, in order to reduce the conservativeness of day-ahead scheduling strategy and improve the economic performance, it is necessary to characterize the uncertainties in the DRO framework. As a result, how to establish the coupling relation of the two networks with game theory in the day-ahead scheduling problem under multiple uncertainties is extremely challenging. Meanwhile, it is difficult to investigate the impact of uncertainties on game-theoretic model and derive the effective method for solving the DRO problem. Table I represents a summary of the literature review and the contributions of this paper. The major contributions of this paper are as follows:

1) We propose a novel day-ahead power scheduling approach for power-transportation coupled network based on game-theory. Specifically, the spatial and temporal distribution models of microgrids (MGs) loads and EV traffic flow in response to the electricity price strategies are established, respectively. A potential function is found to establish the existence of Nash equilibrium and an equivalent formulation of a centralized optimization problem under multiple uncertainties, which is shown that the decision-making is individual optimal.

2) Under the above theoretical framework, the centralized optimization problem can be transformed into a two-stage DRO model. In this way, the master-subproblem framework can be obtained through the duality theory and robust counterpart conversion method, which can be solved by the Benders decomposition approach effectively.

3) The proposed scheduling approach is simulated on the system modified from an IEEE 33-bus system with PVs. Simulations show that the proposed approach has less conservativeness and better economic performance for the day-ahead scheduling profile, which can adapt to the intra-day individual optimal traveling and charging strategies. Meanwhile, the configuration of the power network can improve the flexibility of system energy dispatch, which is closer to the actual application.

The rest of the paper is organized as follows. In Section \uppercase\expandafter{\romannumeral2}, the basic structure and mathematical model of the coupled network are introduced. In Section \uppercase\expandafter{\romannumeral3}, we propose a game-theoretic model with uncertainties for the coupled network and present the equilibrium properties. Section \uppercase\expandafter{\romannumeral4} and \uppercase\expandafter{\romannumeral5} define the ambiguity set and present a DRO two-stage optimization model, and then the corresponding solution methodology is introduced. Numerical results and conclusions are analyzed in Sections \uppercase\expandafter{\romannumeral6} and \uppercase\expandafter{\romannumeral7}, respectively.

\renewcommand\arraystretch{1.5}
\begin{table*}[htbp]
\centering
\caption{Comparison of the proposed model in this paper with recent studies\label{tab:table1}}
\begin{tabular*}{0.9\linewidth}{c|ccc|c|ccc}
\toprule
\multirow{2}*{Ref.} & \qquad  \qquad \qquad \qquad& Netwok Model & \qquad & Multiple & \qquad \qquad \qquad & Method & \qquad \\ 
\quad & TN & PN & Potential Game Relation & Uncertainties & SO & RO & DRO\\
\midrule
\cite{W. Gan_Coordinated} & \checkmark & \checkmark & $\times$ & $\times$  & — & — & —  \\
\cite{S. Bahrami_A Potential}  & \checkmark & \checkmark & \checkmark & $\times$  & — & — & — \\
\cite{Z. Zhou_Power-traffic} & \checkmark & \checkmark & \checkmark & $\times$ & — & — & — \\
\cite{A. Rezaee Jordehi_Two-stage,S. A. Mansouri_A sustainable } & $\times$ & \checkmark & $\times$ & \checkmark & \checkmark & — & — \\
\cite{S. Xie_Two-stage,W. Wei_Robust }  &  \checkmark & \checkmark & $\times$ & \checkmark & — &  \checkmark& — \\
\cite{C. He_Co-optimization,S. Dehghan_Robust} & $\times$ & \checkmark & $\times$ & \checkmark & — &  \checkmark& — \\
\cite{A. Zare_A distributionally,D. Pozo_An ambiguity-averse,Y. Cao_Optimal,C. Wang _ Risk-based,C. Zhao_Data-driven,C. Ning _Data-driven} & $\times$ & \checkmark & $\times$ & \checkmark & — & — &  \checkmark \\
\cite{Y. Zhang_Distributionally}  &  \checkmark & \checkmark & $\times$ & \checkmark & — & — &  \checkmark \\
Our paper  &  \checkmark & \checkmark & \checkmark & \checkmark & — & — &  \checkmark \\
\bottomrule
\end{tabular*}
\end{table*}

\section{Problem  formulation and model}
In this paper, daily 24-hour profiles are employed to represent the dynamic characteristics of active network management. The hour set can be defined as $\mathcal{T}$, $\forall t \in \mathcal{T}$, and $\mathcal{T}=\{1, 2,\cdots, 24\}$.

\subsection{Transportation System Modeling}
Assumptions: Herein, we make some important modeling assumptions for the transportation system. 1) It is a non-atomic measure for each traveler to control a negligible traffic flow and the influence of a single vehicle is infinitesimal. 2) The monetary value of travel time for EV travelers is homogeneous, which is represented by $\omega$. According to Ref. \cite{S. Xie_Two-stage}, it is widely accepted that the value of $\omega$ is constant. 3) The heterogeneous information of vehicles is neglected, such as the state of charging, unit energy consumption, and capacity of the battery. 4) The gasoline vehicles (GVs) and EVs without plenty of electricity to destinations are neglected. 5) We assume that EV travelers can acquire the traffic congestion information for each path and the electricity price of each FCS, and then each traveler chooses the minimum cost route spontaneously to travel and charge. Every EV has to pass through and charge at one FCS before reaching the destination.

A TN can be represented by $\mathcal{G}_{TN}=(\mathcal{N},\mathcal{L})$, where $\mathcal{N}$ and $\mathcal{L}\subseteq\mathcal{N}\times\mathcal{N}$ denote the set of nodes and links, respectively. A pair $(a,b)\in\mathcal{L}$ denotes the link from node \emph a to node \emph b.

1)	\emph {Traffic demand}

In the transportation network, each vehicle has a pair of origin and destination and travels between them, which are named O-D pairs. The set of O-D pairs is denoted by $\mathcal{R}$, specifically, $(o,d)\in\mathcal{R}$. The traffic demand (also called trip rate) of O-D pairs in time slot \emph t can be described by $\boldsymbol{{q}}_{t}^{od}$ (1), which defines the number of vehicles in each time slot intending to travel from \emph o to \emph d.
\begin{equation}
\boldsymbol{q}_{t}^{od}=\begin{bmatrix}
q_{1}^{od},&\cdots,&q_{24}^{od}
\end{bmatrix}^{\rm T}
\end{equation}

2)	\emph {O-D flow}

Each O-D pair is connected by several paths, which consists of certain links \cite{W. Wei_Expansion}. The set of feasible routes \emph p is denoted by $p\in\mathcal{P}_{od}, \mathcal{P}_{od}=\{1,\cdots,N_{od}\}$, where $N_{od}$ denotes the number of the routes for each O-D pair. Let $f_{p,t}^{od}$ be the traffic flow of path \emph p, and the traffic demand balance equation is described by (2). Eq. (3) denotes the non-negativity of path flows.
\begin{equation}
q_{t}^{od}=\sum_{p=1}^{N_{od}}f_{p,t}^{od}
\end{equation}
\begin{equation}
f_{p,t}^{od}\ge0
\end{equation}

3)	\emph {Link flow}

In the transportation system, the set of all links with FCS is represented by $\mathcal{L}_{C}\subseteq\mathcal{L}$. The traffic link flow is equal to the total number of vehicles of the paths through it, which is described by (4).
\begin{equation}
x_{l,t}(f_{p,t}^{od})=\sum_{od\in \mathcal{R}} \sum_{p\in \mathcal{P}_{od}}f_{p,t}^{od}\delta_{l,p}^{od}
\end{equation}

\noindent where an indicator variable $\delta_{l,p}^{od}$ is noted as reflecting the link-path relation. The value of $\delta_{l,p}^{od}$ is set to $1$, if link $l\in\mathcal{L}_p$, where $\mathcal{L}_p$ is a set of links belonging to path $p$; otherwise, $\delta_{l,p}^{od}$ is set to $0$.

In this paper, we assume that FCSs are located on some links. EVs can choose to refill in FCSs or pass by the bypass link. The bus index of distribution systems is marked as $j$, and we define the following constraints to ensure that all EVs are charged through the TN.
\begin{equation}
\sum_{j} x_{j, t}=\sum_{o d} \sum_{p} f_{p, t}^{o d}
\end{equation}
\begin{equation}
f_{p, t}^{o d} \leq \sum_{j} \delta_{j, p}^{o d} x_{j, t}
\end{equation}
\begin{equation}
x_{j, t} \leq \sum_{o d} \sum_{p} \delta_{j, p}^{o d} f_{p, t}^{o d}
\end{equation}

\noindent where $x_{j, t}$ denotes the number of EVs charging at bus $j$, and $\delta_{j, p}^{o d}$ is noted as reflecting the bus-path relation. Specifically, Eq. (5) denotes the balance relationship of the traffic flows of charging link and path. Constraint (6) sets that the traffic flow must be less or equal to the total traffic flows of all links passing the FCS. Constraint (7) denotes that the number of charging EVs for one link must be less or equal to the total traffic flows passed by the link. 

4)	\emph {Traffic expense}

EV travelers choose their driving routes mainly based on the total travel time, which is affected by the congestion level. The travel time can be described by a latency function $t_{l,t}(x_{l,t})$ (8), which is also called the Bureau of Public Roads (BPR) function \cite{Bureau of Public Roads} and can reflect the delayed travel time in links accurately,
\begin{equation}
t_{l,t}(x_{l,t})=t_{l}^{0}\begin{bmatrix}
1+0.15(x_{l.t}/C_{l})^4
\end{bmatrix}\ (l\in \mathcal{L})
\end{equation}
\begin{equation}
x_{l,t}\le C_{l}
\end{equation}

\noindent where $t_{l}^{0}$ is free-speed time and $C_{l}$ is the link flow when $t_{l,t}=1.15t_{l}^{0}$, i.e., the link capacity. For constraint (9), if $x_{l,t}> C_{l}$, the travel time will be penalized by a quick growth. According to (8), the congestion time of each vehicle can be derived by (10).
\begin{equation}
t_{l, t}^{d e}\left(x_{l, t}\right)=t_{l, t}-t_{l}^{0}=0.15 t_{l}^{0}\left(x_{l, t} / C_{l}\right)^{4}
\end{equation}

Furthermore, we assume that the charging demand of each EV in time slot $t$ is uniform, in other words, the power demand of EVs for each bus is proportional to the link flow. $\lambda_{j,t}$ is defined as the electricity price of FCS on bus $j$, and then the charging energy cost of a single vehicle is $\lambda_{j,t}e_{t}$, where $e_{t}$ is the charging demand of one vehicle. It is noted that the charging time of EV is neglected in travel cost due to the same charging efficiency. The total expense by a single EV on path \emph p from \emph o to \emph d in time slot \emph t can be calculated as
\begin{equation}
\operatorname{cost}_{t, p}^{o d}=\sum_{l \in \mathcal{L}} \omega t_{l, t}^{d e}\left(x_{l, t}\right) \delta_{l, p}^{o d}+\sum_{j \in \mathcal{B}} \lambda_{j, t} e_{t} \delta_{j, p}^{o d}
\end{equation}

\noindent where $\omega$ is the monetary value of travel time and $\mathcal{B}$ is the set of buses.

5)	\emph {Wardropian traffic assignment}

The balanced distribution of traffic flow is analyzed by the Wardrop user equilibrium (UE) principle, which can achieve a stable situation of traffic network that no vehicle can decrease the total cost by changing its driving route unilaterally. In other words, a stable situation occurs when total travel costs on all active routes are equal. The UE principle, which is also called Wardrop’s first principle, is closer to the real situation than the second principle \cite{M. J. Beckmann_Studies}.

Then a TAP considering the charging cost is described as follows, in which each traveler wishes to minimize his travel expense.
\begin{equation}
\begin{split}
&{\rm TAP:}\min_{f_{p,t}^{od}}\int_{0}^{x_{l,t}(f_{p,t}^{od})}cost_{t,p}^{od}(\theta)d\theta \\&\text { s.t. } \ (2)-(11)
\end{split} 
\end{equation}

According to classic transportation theory, the UE traffic state can be characterized as (13).
\begin{equation}
\left\{\begin{array}{lr}
f_{p,t}^{od}(cost_{t,p}^{od}-u_{t}^{od})=0,& \\
cost_{t,p}^{od}-u_{t}^{od}\ge 0
\end{array}
\right.
\end{equation}

\noindent where $u_{t}^{od}$ is the minimal travel cost for each vehicle between an O-D pair.

\textbf{Proposition 1:} The solution to problem (12) is an UE flow pattern satisfying (13).

Proof: The proof of this equivalency can be acquired by generalizing the method in \cite{Y. Sheffi_Urban}. In detail, the Lagrangian function of TAP for the equality constraint (2) and inequality constraint (3) can be formulated as
\begin{equation}
\begin{aligned}
L_{\rm TN}(f_{p,t}^{od},u_{t}^{od},&v_{t,p}^{od}) = \int_{0}^{x_{l,t}(f_{p,t}^{od})}cost_{t,p}^{od}(\theta)d\theta+  \\
&\sum_{od\in \mathcal{R}}
\begin{bmatrix}
u_{t}^{od}(q_{t}^{od}-\sum\limits_{p\in \mathcal{P}^{od}}f_{p,t}^{od})-\sum\limits_{p\in \mathcal{P}^{od}}v_{t,p}^{od}f_{p,t}^{od}
\end{bmatrix}
\end{aligned}\notag
\end{equation}
where $u_{t}^{od}$ and $v_{t,p}^{od}$ are the vectors of Lagrangian multipliers. The optimal solution of (12) satisfies the Karush-Kuhn-Tucher conditions as follows:

Stationarity: $\frac{\partial L_{\mathrm{TN}}}{\partial f_{p, t}^{o d}}=0$.

Complementary slackness: $v_{t,p}^{od}f_{p,t}^{od}=0$.

Further considering that
\begin{small}
\begin{equation}
\begin{aligned}
\frac{\partial L_{\mathrm{TN}}}{\partial f_{p, t}^{o d}} &=\omega \sum_{l} t_{l, t}\left(x_{l, t}\right) \frac{\partial x_{l, t}\left(f_{p, t}^{o d}\right)}{\partial f_{p, t}^{o d}}+\sum_{j} \lambda_{j, t} e_{t} \delta_{j, p}^{o d}-u_{t}^{o d}-v_{t, p}^{o d} \\
&=\sum_{l} \omega t_{l, t}^{d e}\left(x_{l, t}\right) \delta_{l, p}^{o d}+\sum_{j} \lambda_{j, t} e_{t} \delta_{j, p}^{o d}-u_{t}^{o d}-v_{t, p}^{o d}\\
&=cost_{p, t}^{o d}-u_{t}^{o d}-v_{t, p}^{o d}\\
&=0
\end{aligned}\notag
\end{equation}
\end{small}

Then $v_{t, p}^{o d}=cost_{t,p}^{od}-u_{t}^{o d}\ge 0$ and $v_{t,p}^{od}f_{p,t}^{od}=0$ can be obtained, which is equivalent to (13).\qed

It is observed that satisfying condition (13) is equivalent to reaching the traffic UE state, and thus it can be considered as a complementary constraint for the optimization problem (12), whose UE state of the optimal solution can be ensured \cite{W. Wei_Quantifying}. Meanwhile, the constraint (13) can be linearized as follows by the big-M method \cite{J. Fortuny-Amat_A representation}.
\begin{equation}
\left\{\begin{array}{lr}
0\le f_{p,t}^{od}\le M(1-w_{p,t}^{od}),& \\
0\le cost_{t,p}^{od}-u_{t}^{od}\le Mw_{p,t}^{od}
\end{array}
\right.
\end{equation}

\noindent where $M$ is a large enough constant and $w_{p,t}^{od}$ is the introduced binary variable, i.e., $w_{p,t}^{od}\in\{0,1\}$.

Furthermore, since the bivariate terms exist in Eq. (10), a piecewise linearization method \cite{S. Xie_Two-stage} is adopted to transform it as (15).
\begin{equation}
\begin{split}
&t_{l,t}^{de}=\sum_{h}^{H}(g_{l,h} \Delta x_{l,h,t}) \\
&x_{l,t}=\sum_{h}^{H} \Delta x_{l,h,t} \\
&0\le \Delta x_{l,h,t}  \le x_{l}^{max}/H
\end{split}
\end{equation}

\noindent where \emph H is the number of linear segments; $g_{l,h}$ is the linear segment slope; and $\Delta x_{l,h,t}$ is the link flow of segment \emph h.

Consequently, the linear optimization problem for TN can be constructed as follows:
\begin{equation}
\begin{aligned}
&\min \ F_{\rm TN}(f_{p,t}^{od}) \triangleq  \sum_{t}(\sum_{l}\omega t_{l,t}^{de}x_{l,t}+\sum_{j}\lambda_{j,t}e_{t}x_{j,t})\\
&\text{ s.t. } \ (2)-(7),(9)-(11),(14),(15)\\
&\qquad{\forall}l \in \mathcal{L},t \in \mathcal{T},od \in \mathcal{R},p \in \mathcal{P}_{od},j \in \mathcal{B}
\end{aligned}
\end{equation}

The objective of the traffic network optimization problem (16) is to find the optimal traffic flow distribution at a certain electricity price.

\subsection{Power System Modeling}
In this study, there are a dispatchable generator (DG), energy storage (ES) unit and photovoltaic (PV) generation unit locating inside each FCS, which can be regarded as a MG \cite{H. Xin_Distributed}. Without loss of generality, we assume that each bus of power distribution network is connected to a MG. Demands of each MG include flexible demand response load and the charging load of EV users, which is depicted in Fig. 1. The application of ES can improve the flexibility of system power dispatch. The operating costs of MGs can be reduced by adjusting the electricity consumption schedule of the flexible demand response loads. Meanwhile, the demand response load can also obtain certain benefits by providing this service. Note that we assume MGs can sell their redundant energy to the main grid. 

In the electrical power network, the index of MGs coupled with transportation network is denoted by $j$, which is also connected on the bus $j\in\mathcal{B}=\{1,2,\cdots,N_{b}\}$, where $N_{b}$ is the number of buses or MGs.
\begin{figure}[htbp]
\centering
\includegraphics[width=3in]{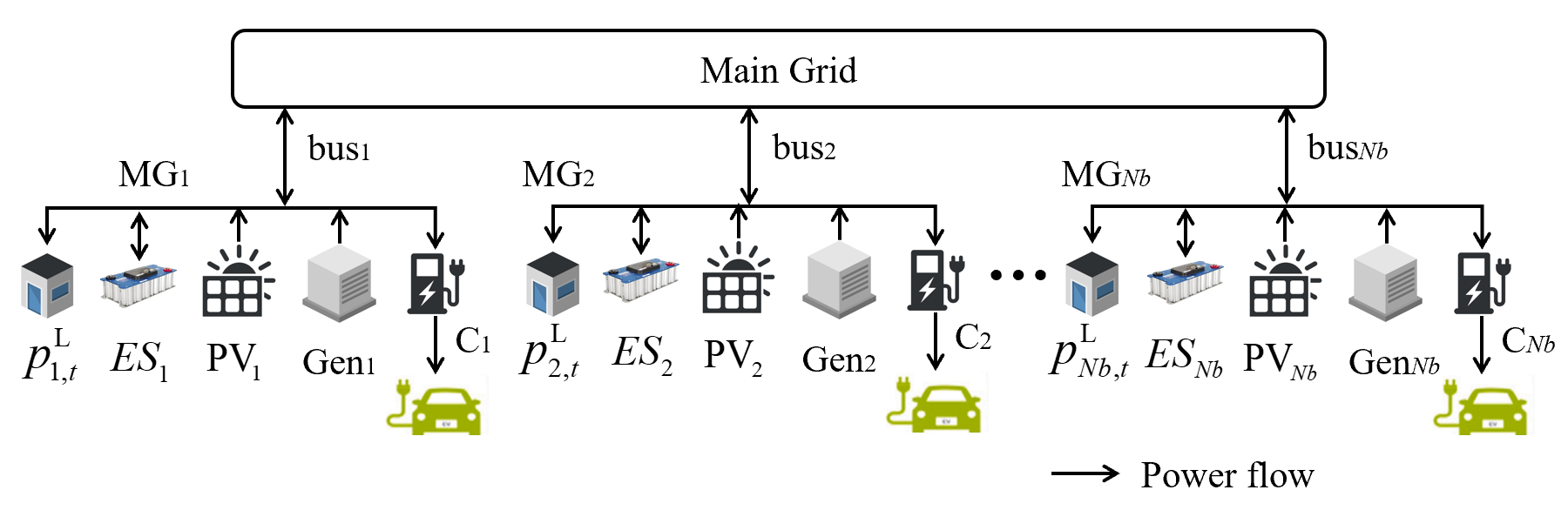}
\caption{Illustration of MG electrical network.}
\label{figl}
 \end{figure}

It should be noted that the adjacent node of bus $j$ is denoted by $i\in\mathcal{N}$, and subscript $ji$ indicates the transmission line from $j$ to $i$. The objective is to maximize the operation revenue of MG $j$. The constraints consist of the following five parts.

1)	\emph {Transaction with the main grid}

It is necessary for MGs to purchase power from the main grid when the power generation units cannot meet the load demand. On the contrary, the MGs can sell the surplus energy of ES or PV generation to the main grid to obtain some revenue. The transaction cost with the main grid in time slot $t$ can be expressed as
\begin{equation}
C_{t}^{\rm G}=\lambda_{t}^{\rm G}\left(p_{j,t}^{\rm buy}-p_{j,t}^{\rm sell}\right)
\end{equation}
\noindent where $p_{j,t}^{\rm buy}$ and $p_{j,t}^{\rm sell}$ are the power purchased and sold by MG in period $t$, respectively. $\lambda_{t}^{\rm G}$ is the day-ahead transaction price of the main grid. The interactive power between MG and main grid must satisfy the constraints as follows.
\begin{equation}
0 \leq p_{j, t}^{\rm buy} \leq u_{j, t} P_{\max }^{\rm{G}}
\end{equation}
\begin{equation}
0 \leq p_{j, t}^{\text {sell }} \leq\left(1-u_{j, t}\right) P_{\max }^{\rm{G}}
\end{equation}
\noindent where $P_{\max }^{\rm G}$ represents the maximum value of the exchange power between the MG and the main grid. It can be influenced by the capacity of the transformer at the connection between the distribution network and the MG. $u_{j, t}$ is a binary variable representing the electricity purchase and sale status of MG $j$. The MG purchases electricity from the main grid when the value of $u_{j, t}$ is $1$, otherwise, the MG sells electricity to the main grid.

2) \emph {The generation constraint of DG}

The power generation cost of DG in MG $j$ can be calculated by Eq. (20).
\begin{equation}
C_{t}^{\rm DG}=a\left(p_{j, t}^{\rm DG}\right)^{2}+b p_{j, t}^{\rm DG}+c
\end{equation}
\noindent where $a$, $b$ and $c$ denote the cost parameters, and $p_{j, t}^{\rm DG}$ is the output power of DG in time slot $t$. The ramp rate constraint is not considered of DG, since that the power response speed is faster than that of hourly scheduling. The output power constraint is considered as following,
\begin{equation}
P_{j, \min }^{\rm DG} \leq p_{j, t}^{\rm DG} \leq P_{j, \max }^{\rm DG}
\end{equation}
\noindent where the feasible range of DG power output is specified. 

3)	\emph {The ES}

The cost of ES can be expressed as the charge and discharge cost by Eq. (22).
\begin{equation}
C_{t}^{\rm ES}=\lambda^{\rm ES}\left(\eta_{C} p_{j, t}^{\rm ES,C}+p_{j, t}^{\rm ES,D} / \eta_{D}\right)
\end{equation}
\noindent where $\lambda^{\rm ES}$ denotes the unit charge and discharge cost, $p_{j, t}^{\rm ES,C}$ and $p_{j, t}^{\rm ES,D}$ represent the charge and discharge power of the ES inverter, respectively. $\eta_{C}$ and $\eta_{D}$ are the charge and discharge efficiency of the ES unit. In addition, the operation constraints of the ES units include functions (23)-(27) as follows.
\begin{equation}
E_{j, t+1}=E_{j, t}+\eta_{C} p_{j, t}^{\rm ES,C} \Delta t-p_{j, t}^{\rm ES,D} \Delta t / \eta_{D}
\end{equation}
\begin{equation}
0 \leq p_{j, t}^{\rm ES,C} \leq v_{j, t} P_{j, \max }^{\rm ES}
\end{equation}
\begin{equation}
0 \leq p_{j, t}^{\rm ES,D} \leq\left(1-v_{j, t}\right) P_{j, \max }^{\rm ES}
\end{equation}
\begin{equation}
E_{j, 0}=E_{j, \rm T}
\end{equation}
\begin{equation}
S O C_{j, \min } \leq E_{j, t} / E_{\rm L} \leq S O C_{j, \max }
\end{equation}

The Eq. (23) represents the remaining capacity of the ES in each period $t$. $\Delta t $ denotes the time interval, whose value is $1h$. Constraints (24) and (25) are the charge and discharge power limits of the ES, respectively. $P_{j, \max }^{\rm ES}$ is the maximum charge and discharge power, which is mainly affected by the capacity limitation of inverters. $v_{j, t}$ represents the charging and discharging state of the ES, The value of $v_{j, t}$ is $1$ when the state of the ES is charging, otherwise, it is $0$. Constraints (26) and (27) describe the $SOC$ limits, where $E_{\rm L}$ is the rated capacity of the ES device. $T$ is the scheduling period, and the value is $24h$. It is noted that the cost of ES equipment life loss is ignored in day-ahead power dispatch.

4)	\emph {The flexible demand response load}

The MG can adjust the power consumption plan of the demand response load flexibly. Meanwhile, users need to be compensated from the MG. The operating cost $C_{t}^{\rm DR}$ in the time slot $t$ can be expressed as
\begin{equation}
C_{t}^{\rm DR}=\lambda_{\rm DR}\left|p_{j, t}^{\rm L}-p_{j, t}^{\rm LE}\right|
\end{equation}
\noindent where $\lambda_{\rm DR}$ represents the unit dispatch cost of demand response load. $p_{j, t}^{\rm LE}$ and $p_{j, t}^{\rm L}$ are the expected power consumption and the actual dispatched power of the MG to the demand response load in time slot $t$, respectively. The auxiliary variable $p_{j, t}^{\rm{L}, \emph{u}}$ and $p_{j, t}^{\rm{L}, \emph{d}}$ can be introduced to substitute the absolute value terms in Eq. (28), which can be replaced by constraints (29)-(31).
\begin{equation}
p_{j, t}^{\rm L}-p_{j, t}^{\rm L E}+p_{j, t}^{\rm{L}, \emph{u}}-p_{j, t}^{\rm{L}, \emph{d}}=0
\end{equation}
\begin{equation}
p_{j, t}^{\rm{L}, \emph{u}}, p_{j, t}^{\rm{L}, \emph{d}} \geq 0
\end{equation}
\begin{equation}
C_{t}^{\rm  DR}=\lambda_{\rm  DR}\left(p_{j, t}^{\rm{L}, \emph{u}}+p_{j, t}^{\rm{L}, \emph{d}}\right)
\end{equation}

The power dispatch of demand response load satisfies the following constraints.
\begin{equation}
P_{j, t}^{\rm L, \min } \leq p_{j, t}^{\rm L} \leq P_{j, t}^{\rm L, \max }
\end{equation}
\begin{equation}
\sum_{t} p_{j, t}^{\rm L} \Delta t=P_{j}^{\rm L}
\end{equation}
\noindent where $P_{j}^{\rm L}$ is the total power demand for one scheduling period, $P_{j, t}^{\rm L, \max }$ and $P_{j, t}^{\rm L, \min }$ denote the maximum and minimum power demand in time slot $t$. 

5)	\emph {The power flow constraint}

For the power flow constraint of this network, the DC approximation of AC optimal power flow is applied to simplify this model. The fixed voltage magnitudes and small phase angle differences are assumed in the network. The reactive part of electrical power is neglected.
\begin{equation}
\sum_{i \in \mathcal{N}(j)} p_{j i, t}=p_{j, t}^{\rm b u y}-p_{j, t}^{\rm s e l l}+p_{j, t}^{\rm P V}+p_{j, t}^{\rm D G}+p_{j, t}^{\rm E S, D}-p_{j, t}^{\rm E S, C}-p_{j, t}^{d}
\end{equation}
\begin{equation}
p_{j, t}^{d}=p_{j, t}^{\rm L}+x_{j, t} e_{t} / \Delta t
\end{equation}
\begin{equation}
p_{j i, t}=b_{j i}\left(\theta_{j}-\theta_{i}\right)
\end{equation}
\begin{equation}
-F_{j i}^{\max } \leq p_{j i, t} \leq F_{j i}^{\max }
\end{equation}

Eq. (34) models the active power flow, which is determined by Kirchhoff’s first law and the power flow equations and $p_{j, t}^{\rm P V}$ is the PV power output. The power on each node has to be balanced. In addition, the electricity demands of MG $j$ include the flexible demand response load and the charging demand, which depend on the traffic flow of the pertaining link. The active power flow on transmission line $ji$ is the multiply by voltage angle (in radians) differences $\theta_{j}-\theta_{i}$ and susceptance $b_{j i}$ as Eq. (36), where $b_{j i}$ is the power distribution factor for line $ji$ and $b_{j i}>0$. For the power flow limits (37), $F_{j i}^{\max }$ is the transmission capacity for line $ji$.

Based on the above five categories constraint, let vector $\boldsymbol{p}_{j, t} \triangleq\left\{p_{j, t}^{\rm b u y}, p_{j, t}^{\text {\rm sell}}, p_{j, t}^{\rm D G}, p_{j, t}^{\rm E S, D}, p_{j, t}^{\rm E S, C}, p_{j, t}^{d}\right\}$ be the decision vector of MG $j$ in time slot $t$. The objective of MG $j$ is to maximize the operation revenue, and the optimization problem is shown as follows. Note that the cost of PV generation is ignored.
\begin{equation}
\begin{aligned}
&\max R_{j}\left(\boldsymbol{p}_{j, t}\right) \triangleq \sum_{t}-\left(C_{t}^{\rm D G}+C_{t}^{\rm E S}+C_{t}^{\rm D R}+C_{t}^{\rm G}\right) \Delta t \\
&\text { s.t. }(17)-(27),(29)-(37) \\
&t \in \mathcal{T}, i, j \in \mathcal{B}
\end{aligned}
\end{equation}

In this power-transportation coupled network, each EV traveler intends to minimize its total travel cost by route selection autonomously under the uncertainty of traffic demand, and the traffic flow distribution can be analyzed by the convex optimization model (16). Meanwhile, each MG operator determines the day-ahead power schedule according to the charging load caused by traffic flows under the uncertainty of PV generation. As a result, the supply and demand of the power network are matched. In addition, the uncertainties of this network will be discussed in Section \uppercase\expandafter{\romannumeral4}.

\section{Game-theoretic analysis}
In this section, we construct the EV charging behavior as a potential game to optimize the costs of MGs and the payoffs of drivers simultaneously. In addition, obtaining a Nash equilibrium of this game is equivalent to solving a centralized optimization problem, which is transformed from Section \uppercase\expandafter{\romannumeral2} \cite{D. Monderer_Potential games}, \cite{Q. La_Potential game}. Based this theory framework, the individual optimality of the coupled network can be ensured.

\subsection{Game-theory with uncertainty}
In the power network, MGs determine the power schedule and maximize their operation revenues by providing energy at the electricity prices. In the transportation network, selecting the lowest-cost routes and charging in FCSs for EV drivers is determined by the electricity prices and the degree of traffic congestion. Meanwhile, the uncertain variables in the two networks are ${p}_{j,t}^{\rm PV}$ and ${q}_{t}^{od}$, respectively. There are interconnections between the strategies and revenues of the EV travelers and MGs, and thus it can be interpreted as a noncooperative game with uncertainties.

\textbf{Definition 1:} The game can be defined by a triplet $\Xi=\Big\{\{\mathcal{Q}\cup\mathcal{B}\},\{\{\mathcal{F}_{v}^{o d}\}_{od\in\mathcal{R}},\{\mathcal{P}_{j}\}_{j\in\mathcal{B}}\},\{(-F_{{\rm TN}v})_{v\in\mathcal{Q}},(R_{j})_{j\in\mathcal{B}}\}\\\Big\}$, and uncertain variables of the coupled network exist in the strategy sets. The components of game $\Xi$ are described as follows:

1)	Customer Side (EVs)

The players noted as $\mathcal{Q}$ in the traffic network are the EV travelers corresponding to the travel demand in different O-D pairs. Player \emph v belongs to $\mathcal{Q}\triangleq[0,{q}_{t}^{od}]$, which is an interval for the number of players. The strategy set consists of  all routes connecting the O-D pair, in other words, the traffic flow on each path can be regarded as the strategy of traffic side, i.e., $\mathcal{F}_{v}^{o d}=\left\{\boldsymbol{f}_{v, 1}^{o d}, \boldsymbol{f}_{v, 2}^{o d}, \cdots, \boldsymbol{f}_{v, N_{o d}}^{o d}\right\}$ , where $\boldsymbol{f}_{v}^{o d}$ is the vector of the traffic flow on the O-D pair in a scheduling cycle. Each EV traveler aims to selfishly choose its route and FCS to maximize its traveling utility with uncertain travel demand, i.e., $
\max\limits _{\boldsymbol{f}_{v}^{o d} \in \mathcal{F}_{v}^{o d}}\left(-\sum\limits _{t} \operatorname{cost}_{t, p}^{o d}\right)
$.

2)	Supplier Side (MGs)

The players noted as $\mathcal{B}$ in the power network are MG operators, player \emph j belongs to $\mathcal{B}\triangleq\{1,2,\cdots,N_{b}\}$, where $N_{b}$ is the number of MGs. The power decision vector for MG $j$ is $\boldsymbol{p}_{j}$ in one scheduling cycle. $\mathcal{P}_{j}=\{\boldsymbol{p}_{1},\boldsymbol{p}_{2},\cdots,\boldsymbol{p}_{N_{j}}\}$ is the strategy set, where $N_{j}$ is the number of MG $j$’s power generation strategies. Then the revenue maximization problem for MG $j$ with uncertain PV generation can be described by $\max\limits _{\boldsymbol{p}_{j}\in\mathcal{P}_{j}}R_{j}$.

The above game-theoretic perspective of this coupled network is shown in Fig. 2, where $\tilde{q}_{t}^{od}$ and $\tilde{p}_{j,t}^{\rm PV}$ denote the uncertain parameters of traffic demand and PV power output. The game is designed to derive the global optimums for EVs and MGs with the constraints of the optimization problem (16) and (38), respectively. It is noted that the number of player \emph v is uncertain due to the uncertainty of traffic demand, which leads to the existence of an uncertain vector in the strategies of the transportation network. Meanwhile, when $q_{t}^{od}$ is determined, $f_{p,t}^{od}$ is variational with the changing of the route selection and can be the decision variable of EV travelers. Note that  the power network revenue function does not contain the uncertain variable $p_{j}^{\rm PV}$, which is only reflected in constraints.

\begin{figure}[htbp]
\centering
\includegraphics[width=3in]{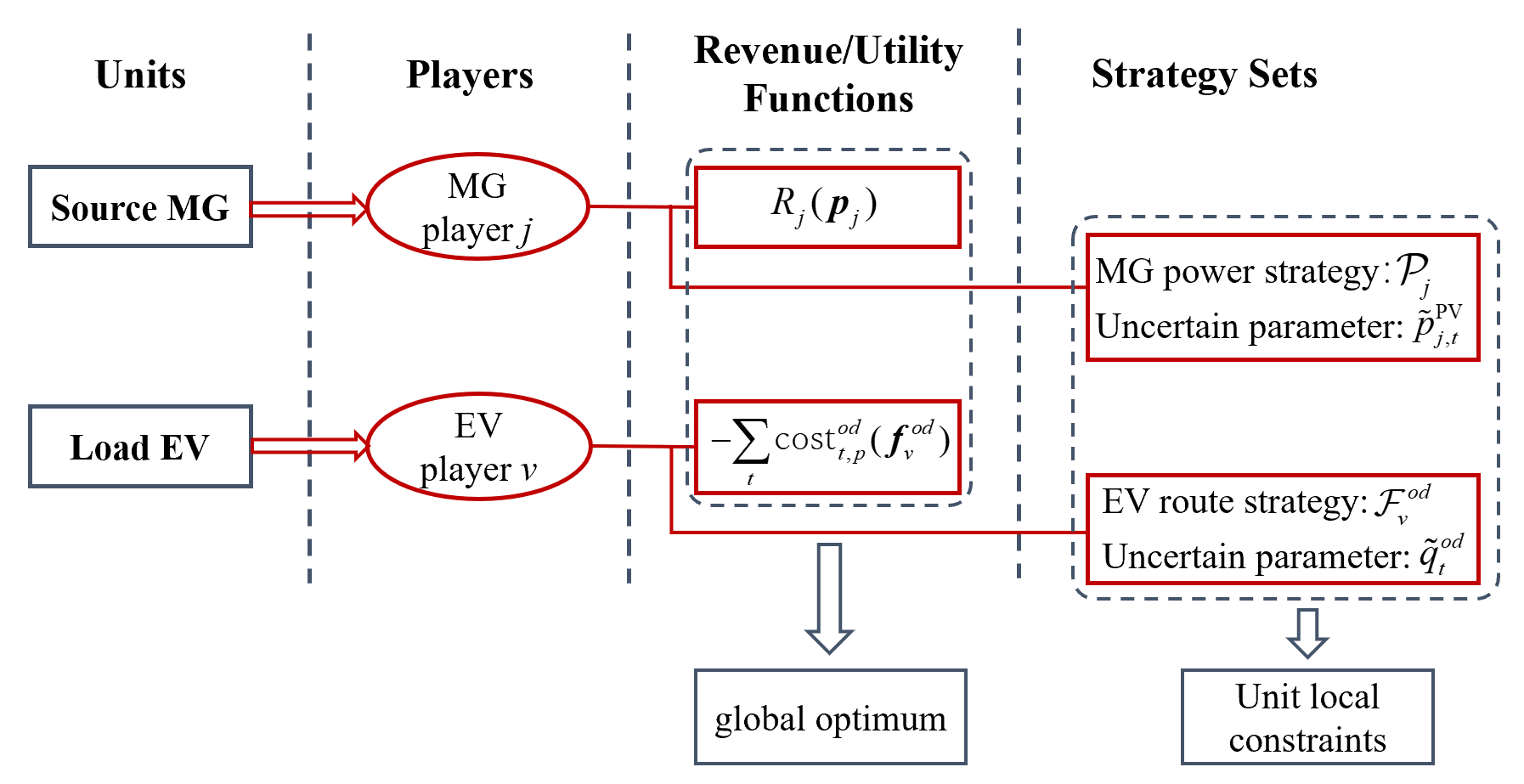}
\caption{The game-theoretic perspective of the power-transportation coupled network.}
\label{figl}
 \end{figure}
 
\subsection{Potential function construction}
Due to the autonomous decision-making of the vehicle, a potential game is formulated. Under the above mathematical framework, the whole problem turns out to be seeking the Nash equilibrium of the game. The corresponding Nash equilibrium is a strategy profile on which no traveler can improve its utility by unilaterally changing its route at a certain electricity price, i.e., the Wardrop UE, and no MG can benefit more by switching to another strategy except the optimal power strategy.

\textbf{Proposition 2:} The Nash equilibrium can be obtained by solving an equivalent centralized optimization problem, and the game is a potential game with potential function as follows:
\begin{equation}
\begin{aligned}
\Phi\left(f_{p, t}^{o d}, \boldsymbol{p}_{j, t}\right)=-\sum_{t}[\sum_{l} \omega t_{l, t}^{d e} x_{l, t}\left(f_{p, t}^{o d}\right)+\sum_{j}(C_{t}^{\rm D G} \\
\phantom{=\;\;}
+C_{t}^{\rm E S}+C_{t}^{\rm D R}+C_{t}^{\rm G}) \Delta t]
\end{aligned}
\end{equation}

\noindent where the variables satisfy the constraints of the optimization problems (16) and (38).

Proof:

1)	Variable Declaration

There are two kinds of decision vectors as follows. In transportation network, $\boldsymbol{f}_{v}^{od}=(f_{v,1}^{od},f_{v,2}^{od},\cdots,f_{v,24}^{od})$ is the traffic flow in the route selected by EV $v$. $\boldsymbol{f}_{-v}^{od}$ is the vector of the traffic flow in the route selected by all EVs except EV $v$, defined as $\boldsymbol{f}_{-v}^{od}=(\boldsymbol{f}_{v}^{od},\cdots,\boldsymbol{f}_{v-1}^{od},\boldsymbol{f}_{v+1}^{od},\cdots,\boldsymbol{f}_{q_{s}^{od}}^{od})$. $(\boldsymbol{f}_{-v},\boldsymbol{f}_{v}^{od})$ and $(\boldsymbol{f}_{-v}^{od},\hat{\boldsymbol{f}}_{v}^{od})$ are the arbitrary two strategies of EV $v$. In the same way, in the power network, $\boldsymbol{p}_{i}$ is the power profile of MG $i$. $\boldsymbol{p}_{-i}$ is the power profiles of all MGs except MG $i$ defined as $\boldsymbol{p}_{-i}=(\boldsymbol{p}_{1},\cdots,\boldsymbol{p}_{i-1},\boldsymbol{p}_{i+1},\cdots,\boldsymbol{p}_{N_{b}})$. $(\boldsymbol{p}_{-i},\boldsymbol{p}_{i})$ and $(\boldsymbol{p}_{-i},\hat{\boldsymbol{p}}_{i})$ are the arbitrary two strategies of MG $i$.

2)	Transportation Network Side

We first let $L_{\mathrm{T}-\mathrm{P}}$ be the partial Lagrangian of $-\mit \Phi$ associated with constraints (2) and (34),which are shown in Eq. (40). It should be noted that the charging price is determined by the marginal cost of power production of FCSs, viz. the dual variables of power balance constraints (34).
\newcounter{TempEqCnt}
\setcounter{TempEqCnt}{\value{equation}}
\setcounter{equation}{39}
\begin{figure*}[ht]
\hrulefill
\begin{equation}
\begin{aligned}
&L_{\mathrm{T}- \mathrm{P}}\left(f_{p, t}^{o d}, \boldsymbol{p}_{j, t}, u_{t}^{o d}, \lambda_{j, t}\right)=\sum_{t}\left[\sum_{l} \omega \mathrm{t}_{l, t}^{d e} x_{l, t}\left(f_{p, t}^{o d}\right)+\sum_{j}\left(C_{t}^{\rm D G}+C_{t}^{\rm E S}+C_{t}^{\rm D R}+C_{t}^{\rm G}\right) \Delta t\right]+ \\
&\sum_{t} \sum_{o d}\left[u_{t}^{o d}\left(\sum_{p} f_{p, t}^{o d}-q_{t}^{o d}\right)\right]-\sum_{t} \sum_{j} \lambda_{j, t}\left[p_{j, t}^{\rm b u y}-p_{j, t}^{\rm s e l l}+p_{j, t}^{\rm P V}+p_{j, t}^{\rm D G}+p_{j, t}^{\rm E S, D}-p_{j, t}^{\rm E S, C}-\left(p_{j, t}^{\rm L}+x_{j, t} e_{t} / \Delta t\right)-\sum_{i \in \mathcal{N}(j)} p_{j i, t}\right]
\end{aligned}
\end{equation}
\hrulefill
\end{figure*}

The Lagrangian for traffic side can be expressed as
\begin{equation}
\begin{aligned}
L_{\rm T}\left(f_{p, t}^{o d}, u_{t}^{o d}\right)=\sum_{t}\left[\sum_{l} \omega t_{l, t}^{d e} x_{l, t}\left(f_{p, t}^{o d}\right)+\sum_{j} \lambda_{j, t} e_{t} x_{l, t}\left(f_{p, t}^{o d}\right)\right] \\
\phantom{=\;\;}
+\sum_{t} \sum_{o d}\left[u_{t}^{o d}\left(\sum_{p} f_{p, t}^{o d}-q_{t}^{o d}\right)\right]
\end{aligned}
\end{equation}
By calculating the Lagrangian partial derivative, (40) and (41) yields
\begin{equation}
\frac{\partial L_{\mathrm{T}-\mathrm{P}}}{\partial f_{p, t}^{o d}}=\frac{\partial F_\mathrm{T N}}{\partial f_{p, t}^{o d}}+u_{t}^{o d}=\frac{\partial L_{\mathrm{T}}}{\partial f_{p, t}^{o d}}
\end{equation}

In addition, let $\boldsymbol{f}^{od}=(f_{1,1}^{od},\cdots,f_{i,24}^{od},\cdots,f_{q_{t}^{od},1}^{od},\\
\cdots,f_{q_{t}^{od},24}^{od})$ denote the travel flow profile vector of all EVs in the O-D pair. Considering the definition of ordinal potential game \cite{Q. La_Potential game}, for two strategy profiles $\boldsymbol{f}^{od}=(\boldsymbol{f}_{-v}^{od},\boldsymbol{f}_{v}^{od})$ and $\hat{\boldsymbol{f}^{od}}=(\boldsymbol{f}_{-v}^{od},\hat{\boldsymbol{f}}_{v}^{od})$, if $[-F_{{\rm TN}_{v}}(\boldsymbol{f}_{-v}^{od},\boldsymbol{f}_{v}^{od})]-[-F_{{\rm TN}_{v}}(\boldsymbol{f}_{-v}^{od},\hat{\boldsymbol{f}_{v}^{od}})]\ge0$, then $\mit \Phi(\boldsymbol{f}_{-v}^{od},\boldsymbol{f}_{v}^{od},\boldsymbol{p}_{i})-\mit \Phi(\boldsymbol{f}_{-v}^{od},\hat{\boldsymbol{f}_{v}^{od}},\boldsymbol{p}_{i})\ge$$0$ can be derived due to the monotonicity and non-negativity of $F_{\rm TN}$. Thus $\mit \Phi$ is a potential function for EV travelers.

3)	Power network Side

In the power network side, the Lagrangian can be expressed as 
\begin{equation}
\begin{aligned}
&L_{\rm P}\left(\boldsymbol{p}_{j, t}, \lambda_{j, t}\right)=\sum_{t} \sum_{j}\left(C_{t}^{\rm D G}+C_{t}^{\rm E S}+C_{t}^{\rm D R}+C_{t}^{\rm G}\right) \Delta t- \\
&\sum_{t} \sum_{j} \lambda_{j, t} [p_{j, t}^{\rm b u y}-p_{j, t}^{\rm s e l l}+p_{j, t}^{\rm P V}+p_{j, t}^{\rm D G}+p_{j, t}^{\rm E S, D}-p_{j, t}^{\rm E S, C}\\
&-\left(p_{j, t}^{\rm L}+x_{j, t} e_{t} / \Delta t\right)-\sum_{i \in \mathcal{N}(j)} p_{j i, t}]
\end{aligned}
\end{equation}

The necessary and sufficient condition for potential function being obtained and the objective function of power network getting the minimum value simultaneously can be expressed as (44), which can be satisfied obviously.
\begin{equation}
\frac{\partial L_{\mathrm{T} - \mathrm{P}}}{\partial \boldsymbol{p}_{j, t}}=\frac{\partial L_\mathrm{P}}{\partial \boldsymbol{p}_{j, t}}
\end{equation}

In addition, for two strategy profiles $\boldsymbol{p}=(\boldsymbol{p}_{-i},\boldsymbol{p}_{i})$ and $\hat{\boldsymbol{p}}=(\boldsymbol{p}_{-i},\hat{\boldsymbol{p}}_{i})$, if $R_{i}(\boldsymbol{p}_{-i},\boldsymbol{p}_{i})-R_{i}(\boldsymbol{p}_{-i},\hat{\boldsymbol{p}_{i}})\ge$$0$, it can be proofed readily that $\mit{\Phi}(\boldsymbol{p}_{-i},\boldsymbol{p}_{i},\boldsymbol{f}_{v}^{od})-\mit{\Phi}(\boldsymbol{p}_{-i},\hat{\boldsymbol{p}_{i}},\boldsymbol{f}_{v}^{od})\ge$$0$. Thus $\mit \Phi$ is a potential function for MGs. \qed

In the potential game $\Xi$, the utility and revenue function for each player can be mapped to the potential function $\mit{\Phi}(f_{p,t}^{od},\boldsymbol{p}_{j,t})$. The Nash equilibrium of game $\Xi$ is equivalent to the set of optimal solutions of the potential function (39), where the feasible set of decision variables can be defined by constraints of the optimization problems (16) and (38). Note that the potential function is concave with respect to decision variables, which guarantees the existence and uniqueness of the Nash equilibrium. Therefore, we can determine the optimal strategy of the coupled network by locating the local optima of the above potential function \cite{D. Monderer_Potential games}.

\section{Ambiguity set and reformulation of DRO problem}
The exogenous uncertainties of the power-transportation coupled network include the PV power output and charging demand of FCS which is led by the uncertainty of the traffic demand. Furthermore, these uncertainties can be transformed into the source-load uncertainties of the power network, which are built as moment-based ambiguity sets incorporating the probability distribution information to reduce the conservativeness by RO in this study. In this framework, the uncertainties of the coupled network can be captured by a two-stage DRO model, in which all game players seek to optimize their profits. Generally, the decision variables in the DRO model include two-stage strategies. The first-stage strategies are mainly the design strategies, such as price setting and power purchase plan in the day-ahead, which must be made “here-and-now” before the realization of uncertainties. The second-stage strategies are the operational decisions and can also be called a “wait-and-see” pattern, which is determined after the uncertainty realization of the coupled network.

\subsection{Uncertainty Handling}
The uncertainty of traffic demand for each O-D pair in time slot $t$ can be characterized as a box uncertainty set \cite{W. Wei_Robust} in (45).
\begin{equation}
\tilde{q}_{t}^{od} \in {\rm BOX}(\underline{q}_{t}^{od},\overline{q}_{t}^{od}) \triangleq \{ \underline{q}_{t}^{od} \le \tilde{q}_{t}^{od} \le \overline{q}_{t}^{od}, \forall od \in \mathcal{R} \}
\end{equation}

The traffic flow space-time distribution led by the uncertainty of traffic demand can be transformed into uncertainties of the charging demand of FCSs connected to the buses, which is illustrated in Fig. 3.
\begin{figure}[htbp]
\centering
\includegraphics[width=3in]{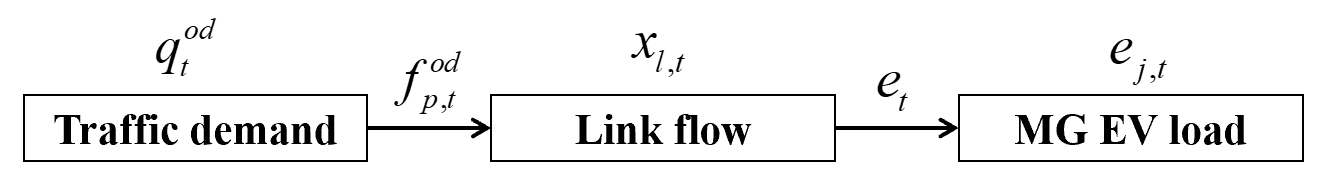}
\caption{Illustration of uncertain variables conversion in a transportation system.}
\label{figl}
 \end{figure}

According to the Wardrop UE Principle, there is a unique solution for the convex optimization problem (12), and the box uncertainty set $\tilde{f}_{p,t}^{od} \in {\rm BOX}(\underline{f}_{p,t}^{od},\overline{f}_{p,t}^{od})$ can be obtained by \cite{W. Wei_Robust}
\begin{equation}
\begin{split}
&\overline{f}_{p,t}^{od}=\left\{\begin{array}{lr}
\mathop{\arg\min} \int_{0}^{x_{l,t}(f_{p,t}^{od})}\rm cost_{t,p}^{od}(\theta)d\theta \\
\text{s.t.} \ \overline{q}_{t}^{od}=\sum\limits_{p\in \mathcal{P}^{od}}f_{p,t}^{od},(3)-(11)
\end{array}
\right.\\
&\underline{f}_{p,t}^{od}=\left\{\begin{array}{lr}
\mathop{\arg\min} \int_{0}^{x_{l,t}(f_{p,t}^{od})}\rm cost_{t,p}^{od}(\theta)d\theta \\
\text{s.t.} \ \underline{q}_{t}^{od}=\sum\limits_{p\in \mathcal{P}^{od}}f_{p,t}^{od},(3)-(11)
\end{array}
\right.
\end{split}
\end{equation}

The link flow is monotonically increasing for the path flow according to Eq. (5). Then the box set of the link flow can be derived as

\begin{equation}
\tilde{x}_{j,t} \in {\rm BOX}(x_{j,t}(\underline{f}_{p,t}^{od}) , x_{j,t}(\overline{f}_{p,t}^{od}))
\end{equation}

Due to the charging demand for each FCS being regarded as a linear relation with the link flow, the uncertainty box set of charging load in MG $j$ can be represented as

\begin{equation}
\tilde{e}_{j,t} \in {\rm BOX}(x_{j,t}(\underline{f}_{p,t}^{od})e_{t}, x_{j,t}(\overline{f}_{p,t}^{od})e_{t})
\end{equation}

In addition, it is assumed that the available energy of PV generation fluctuates with the interval between $\underline{p}_{j,t}^{\rm PV}$ and $\overline{p}_{j,t}^{\rm PV}$ , and then the uncertainty set of PV power output can also be characterized by a box set as follows:
\begin{equation}
\tilde{p}_{j,t}^{\rm PV} \in {\rm BOX}(\underline{p}_{j,t}^{\rm PV}, \overline{p}_{j,t}^{\rm PV})
\end{equation}

\subsection{Ambiguity Set Construction}
There are two common methods to establish an ambiguity set generally, including moment-based \cite{D. Pozo_An ambiguity-averse} and distance metric approach \cite{P. M. Esfahani_Data-driven}. In this study, the moment-based method is adopted to construct a ambiguity set considering the mean and support information.

According to Ref. \cite{Y. Zhang_Distributionally}, the uncertain variable $\tilde{e}_{j,t}$ and $\tilde{p}_{j,t}^{\rm PV}$ can be written as
\begin{equation}
\left\{\begin{array}{lr}
\tilde{e}_{j,t}=e_{j,t,pr}+e_{j,t,de} \tilde{\alpha}_{j,t} \\
\tilde{p}_{j,t}^{\rm PV}=p_{j,t,pr}^{\rm PV}+p_{j,t,de}^{\rm PV} \tilde{\beta}_{j,t}
\end{array}
\right.
\end{equation}

\noindent where $e_{j,t,pr}$ and $p_{j,t,pr}^{\rm PV}$ are the predicted charging demand and predicted PV generation. Parameters $e_{j,t,de}$ and $p_{j,t,de}^{\rm PV}$ denote the maximum deviations relative to the predicted values, satisfying Eq. (51). Random variables $\tilde{\alpha}_{j,t}$ and $\tilde{\beta}_{j,t}$ take values within $[-1,1]$ indicating the degree of fluctuation relative to the predicted values. We assume that $\boldsymbol{\tilde{\alpha}}_{t}=[\tilde{\alpha}_{1,t},\tilde{\alpha}_{2,t},\cdots,\tilde{\alpha}_{j,t},\cdots,\tilde{\alpha}_{N_{b},t}] \in\mathbb{R}^{N_{b}}$ and $\boldsymbol{\tilde{\beta}}_{t}=[\tilde{\beta}_{1,t},\tilde{\beta}_{2,t},\cdots,\tilde{\beta}_{j,t},\cdots,\tilde{\beta}_{N_{b},t}] \in\mathbb{R}^{N_{b}}$ are variable vectors, and then the uncertain vector $\boldsymbol{\tilde{\sigma}}_{t}$ is defined as $\boldsymbol{\tilde{\sigma}}_{t}=\{\boldsymbol{\tilde{\alpha}}_{t}^{\rm T}\ \boldsymbol{\tilde{\beta}}_{t}^{\rm T}\}_{(2N_{b})\times1}^{\rm T}$$(t\in \mathcal{T})$.
\begin{equation}
\begin{aligned}
&e_{j,t,de}=\max\left\{e_{j,t,pr}-x_{j,t}(\underline{f}_{p,t}^{od})e_{t},x_{j,t}(\overline{f}_{p,t}^{od})e_{t}-e_{j,t,pr}\right\}\\
&p_{j,t,de}^{\rm PV}=\max\left\{p_{j,t,pr}^{\rm PV}-\underline{p}_{j,t}^{\rm PV}, \overline{p}_{j,t}^{\rm PV}-p_{j,t,pr}^{\rm PV}\right\}
\end{aligned}
\end{equation}

In this way, the ambiguity set $\mathscr{P}_{t}$ is defined as \cite{W. Wiesemann_Distributionally}
\begin{equation}
\begin{aligned}
\mathscr{P}_{t}=\left\{\mathbb{P}_{t} \in\mathbb{R}^{N_{b}} \times \mathbb{R}^{N_{b}}: \mathbb{E}_{\mathbb{P}_{t}}\left[\tilde{\boldsymbol{\sigma}}_{t}\right]=\mathbf{0}, \right. \\
\phantom{=\;\;}
\left.\mathbb{P}_{t}\left[\tilde{\boldsymbol{\sigma}}_{t} \in \mathscr{C}_{m, t}\right]=P_{m, t}, \forall m \in \mathcal{M}\right\}
\end{aligned}
\end{equation}

\noindent where $\mathscr{P}_{t}$ denotes that the joint probability distribution on $\mathbb{R}^{N_{b}} \times \mathbb{R}^{N_{b}}$ of uncertain vector $\boldsymbol{\tilde{\alpha}}_{t}$ and $\boldsymbol{\tilde{\beta}}_{t}$ is $\mathbb{P}_{t}$, which is obtained by the historical data of the PV power output and the traffic demand. The first item in (52) indicates that the expectation of the uncertain vector in time slot $t$ is equal to $\boldsymbol{0}$. The second item implies that the probability of confidence set $\mathscr{C}_{m, t}$ occurring is $P_{m, t}$ with $\mathcal{M}=\{1,\cdots,M_{0}\}$, and $M_{0}$ is the number of confidence sets. Meanwhile, the value of $P_{m, t}$ belongs to $[0,1]$ for all $m \in \mathcal{M}$ and $P_{M_{0}, t}=1$ is assumed. 

According to the Ref. \cite{W. Wiesemann_Distributionally}, the confidence set $\mathscr{C}_{m, t}$ is defined as
\begin{equation}
\begin{aligned}
\mathscr{C}_{m, t}=\left\{\tilde{\boldsymbol{\sigma}}_{t} \in \mathbb{R}^{N_{b}} \times \mathbb{R}^{N_{b}}:\left\|\boldsymbol{\sigma}_{t}\right\|_{\infty} \leq 1,\left\|\boldsymbol{\sigma}_{t}\right\|_{1} \leq \Gamma_{m, t},\right. \\
\phantom{=\;\;}
\left.\forall t \in \mathcal{T}, \forall m \in \mathcal{M}\right\}
\end{aligned}
\end{equation}
where the first item denotes that all the elements in $\boldsymbol{\tilde{\sigma}}_{t}$ take values between -1 and 1, and the second item implies that the sum of absolute values for all components in $\boldsymbol{\tilde{\sigma}}_{t}$ take values lower than the budget of uncertainty defined as $\Gamma_{m, t}$, which is added to avoid that $\boldsymbol{\tilde{\sigma}}_{t}$ always achieves boundary values to reduce the conservativeness of this model. For obtaining tractable DRO, we assume that $\Gamma_{m, t}$ is strictly increasing with the increase of $m$ and $\mathscr{C}_{m, t}$$\subset$$\mathscr{C}_{m+1, t}$. Furthermore, the probabilities in (52) satisfy ${P}_{m, t}$$\le$${P}_{m+1, t}$ for $\forall m \in\{1,\cdots,$$M_{0}$$-1\}$. It is noted that if $\Gamma_{m, t}$ is given, ${P}_{m, t}$ is the probability for the uncertain vector $\boldsymbol{\tilde{\sigma}}_{t}$ occurring in $\mathscr{C}_{m, t}$.

\subsection{Reformulation of the DRO Problem}
The purpose of this study is to optimize the cost of power generation from the perspective of distribution network and MG, and to achieve the optimal social benefit simultaneously, viz. to ensure the optimality of traffic scheduling. Under the potential game-theory framework, the centralized optimization method is employed to carry out the day-ahead power scheduling and ensure the individual optimality. Due to the existence of uncertain variables, the Nash equilibrium state of potential function (39) can be derived by an equivalent optimization problem as follows:
\begin{equation}
\begin{aligned}
\min \sup _{\mathbb{P}_{t} \in \mathscr{P}_{t}} \mathbb{E}_{\mathbb{P}_{t}}\{\sum_{t \in T} \sum_{l \in \mathcal{L}} \omega t_{l, t}^{d e} x_{l, t}\left(f_{p, t}^{o d}\right)+\sum_{t \in T} \sum_{j \in B}[\lambda_{t}^{\rm G}(p_{j, t}^{\rm b u y}-\\
\phantom{=\;\;}
p_{j, t}^{\rm s e ll})+a\left(p_{j, t}^{\rm D G}\right)^{2}+b p_{j, t}^{\rm D G}+c+\lambda^{\rm E S}(\eta_{C} p_{j, t}^{\rm E S, C}\\
\phantom{=\;\;}
+p_{j, t}^{\rm E S, D} / \eta_{D})+\lambda_{\rm D R}\left|p_{j, t}^{\rm L}-p_{j, t}^{\rm L E}\right|] \Delta t\}\\
\phantom{=\;\;}
\text{s.t.} (2)-(7), (9)-(11), (14), (15), (17)-(27), (29)-(37)
\end{aligned}
\end{equation}

In the following, a two-stage distributionally robust optimization model is proposed with the underlying compact matrix form:
\begin{equation}
\begin{aligned}
&\min _{\boldsymbol{x}} \boldsymbol{c}^{\mathrm{T}} \boldsymbol{x}+\min _{\boldsymbol{y}_{t}} \sum_{t \in \mathcal{T}} \sup _{\mathbb{P}_{t} \in \mathscr{P}_{t}} \mathbb{E}_{\mathbb{P}_{t}}(\boldsymbol{b}^{\mathrm{T}} \boldsymbol{y}_{t})\\
&\text { s.t. } \boldsymbol{A x} \leq \boldsymbol{h} \\
&\qquad \boldsymbol{B}_{t} \boldsymbol{x}+\boldsymbol{C}_{t} \boldsymbol{y}_{t} \leq \boldsymbol{d}_{t}+\boldsymbol{D}_{t} \boldsymbol{\sigma}_{t}
\end{aligned}
\end{equation}

\noindent where $\boldsymbol{A}$, $\boldsymbol{B}_{t}$, $\boldsymbol{C}_{t}$, and $\boldsymbol{D}_{t}$ denote constant matrices; $\boldsymbol{h}$ and $\boldsymbol{d}_{t}$ are constant vectors. Vectors $\boldsymbol{x}$ and $\boldsymbol{y}_{t}$ represent decision variables and are listed as follows:
\begin{equation}
\begin{aligned}
&\boldsymbol{x}=\left\{u_{j, t}, v_{j, t}, p_{j, t}^{\rm b u y}\right\} \\
&\boldsymbol{y}_{t}=\left\{x_{l, t}, x_{j, t}, f_{p, t}^{o d}, p_{j, t}^{\rm s e l l}, p_{j, t}^{\rm D G}, p_{j, t}^{\rm E S, D}, p_{j, t}^{\rm E S, C}, p_{j, t}^{\rm L}\right\}
\end{aligned}
\end{equation}
The variable $\boldsymbol{x}$ is a vector of the first-stage decision consisting of  the variables determined in the day-ahead and the indicator variables. The variable $\boldsymbol{y}_{t}$ is a vector of the second-stage decision, which includes link flow, path flow, and decision variables of power network.

From the objective of (55), the second item differentiates with classical robust approaches, and the physical meaning of this item is to minimize the worst situation depending on the distribution $\mathbb{P}_{t}$. Combined with (52), the second item in the objective of (55) can be converted as follows {\cite{H. Xin_Distributed}:
\begin{equation}
\begin{aligned}
\min _{\boldsymbol{y}_{t}} \sup _{\mathbb{P}_{t} \in \mathscr{P}_{t}} \mathbb{E}_{\mathbb{P}_{t}}(\boldsymbol{b}^{\mathrm{T}} \boldsymbol{y}_{t})=&\sup _{\mathbb{P}_{t} \in \mathscr{P}_{t}} \mathbb{E}_{\mathbb{P}_{t}}(\min \boldsymbol{b}^{\mathrm{T}} \boldsymbol{y}_{t})=\\&
\max \int\limits_{\mathscr{C}_{M_{0}, t}}(\min \boldsymbol{b}^{\mathrm{T}} \boldsymbol{y}_{t}) d P\left(\boldsymbol{\sigma}_{t}\right)
\end{aligned}
\end{equation}

\noindent Similarly, constraints in the ambiguity set (52) about the probability of $\tilde{\boldsymbol{\sigma}}_{t}$ can be transformed as
\begin{equation}
\int_{\mathscr{C}_{M_{0}, t}} \boldsymbol{\sigma}_{t} d P\left(\boldsymbol{\sigma}_{t}\right)=\mathbf{0},\ \int_{\mathscr{C}_{m, t}} d P\left(\boldsymbol{\sigma}_{t}\right)=P_{m, t}
\end{equation}

As a result, we can reformulate (55) as follows \cite{M. Zugno_A Robust}:

\begin{equation}
\begin{aligned}
\min _{\boldsymbol{x}} \boldsymbol{c}^{\mathrm{T}} \boldsymbol{x}+\min _{\boldsymbol{y}_{t}} \sum_{t \in \mathcal{T}} &\sup _{\mathbb{P}_{t} \in \mathscr{P}_{t}} \mathbb{E}_{\mathbb{P}_{t}}(\boldsymbol{b}^{\mathrm{T}} \boldsymbol{y}_{t})\\
&\text { s.t. } \int\limits_{\mathscr{C}_{M_{0}, t}} \boldsymbol{\sigma}_{t} d P\left(\boldsymbol{\sigma}_{t}\right)=\mathbf{0}:\ \boldsymbol{\eta}_{t}\\
&\qquad \int\limits_{\mathscr{C}_{m, t}} d P\left(\boldsymbol{\sigma}_{t}\right)=P_{m, t}:\ \gamma_{m t}
\end{aligned}\notag
\end{equation}
\begin{equation}
\boldsymbol{B}_{t} \boldsymbol{x}+\boldsymbol{C}_{t} \boldsymbol{y}_{t} \leq \boldsymbol{d}_{t}+\boldsymbol{D}_{t} \boldsymbol{\sigma}_{t}:\ \boldsymbol{\nu}_{t}\!\!\!\!\!\!\!\!\!\!\notag
\end{equation}
\begin{equation}
\boldsymbol{A x} \leq \boldsymbol{h}~~~~~~~~~~~~~~~~~~~~~~~~~~~~~~~~~~~~~~~~~~~~~~~~~~~~~~\notag
\end{equation}

\noindent where $\boldsymbol{\eta}_{t}$, $\boldsymbol{\nu}_{t}$ and $\gamma_{m t}$ denote the dual vectors for the corresponding constraints.

\section{Solution Methodology}
Generally, the two-stage optimization problem (55) can be derived as a master problem (MP) and a subproblem (SP). It should be noted that the term including the uncertainties in (55) has a minimax form, $\sup\limits _{\mathbb{P}_{t} \in \mathscr{P}_{t}} \mathbb{E}_{\mathbb{P}_{t}}(\min \boldsymbol{b}^{\mathrm{T}} \boldsymbol{y}_{t})$, which is classified as a bi-level programming problem. For solving the above problem, duality theory can be used to transform the bi-level problem into a single-level model with corresponding dual variables.

\textbf{Proposition 3:} Concerning the ambiguity set $\mathscr{P}_{t}$, the bilevel programming problem $\sup\limits _{\mathbb{P}_{t} \in \mathscr{P}_{t}} \mathbb{E}_{\mathbb{P}_{t}}(\min \boldsymbol{b}^{\mathrm{T}} \boldsymbol{y}_{t})$ can be reformulated as the following optimization problem:
\begin{small}
\begin{equation}
\begin{aligned}
&\min \sum_{m=0}^{M_{0}} \gamma_{m t} P_{m, t} \\
&\text {s.t.}\sum_{s=m-1}^{M_{0}} \gamma_{s t} \geq \\
&\qquad \max\limits_{\boldsymbol{\sigma}_{m t},\boldsymbol{\nu}_{m t}} \left[-(\boldsymbol{\sigma}_{m t})^{\mathrm{T}} \boldsymbol{\eta}_{t}+\left(\boldsymbol{\nu}_{m t}\right)^{\mathrm{T}}\left(\boldsymbol{B}_{t} \boldsymbol{x}-\boldsymbol{d}_{t}-\boldsymbol{D}_{t} \boldsymbol{\sigma}_{m t}\right)\right] \\
&\qquad \boldsymbol{b}=\left(\boldsymbol{C}_{t}\right)^{\mathrm{T}} \boldsymbol{\nu}_{t}
\end{aligned}
\end{equation}
\end{small}

Proof: Concerning the confidence set $\mathscr{C}_{m, t}$$\subset$$\mathscr{C}_{m+1, t}$ for all $m \in \mathcal{M}$,  the objective (57) can be discretized into \cite{Y. Zhang_Distributionally}
\begin{small}
\begin{equation}
\begin{aligned}
&\max \int\limits_{\mathscr{C}_{M_{0}, t}}(\min \boldsymbol{b}^{\mathrm{T}} \boldsymbol{y}_{t}) d P\left(\boldsymbol{\sigma}_{t}\right)\\
&=\max \left[\int\limits_{\mathscr{C}_{1, t}}(\min \boldsymbol{b}^{\mathrm{T}} \boldsymbol{y}_{t}) d P\left(\boldsymbol{\sigma}_{t}\right)+\int\limits_{\mathscr{C}_{2, t} \backslash \mathscr{C}_{1, t}}(\min \boldsymbol{b}^{\mathrm{T}} \boldsymbol{y}_{t}) d P\left(\boldsymbol{\sigma}_{t}\right)
\right.
\\
&\phantom{=\;\;}
\left.+\cdots + \int\limits_{\mathscr{C}_{M_{0,t}} \backslash \mathscr{C}_{M_{0}-1, t}}(\min \boldsymbol{b}^{\mathrm{T}} \boldsymbol{y}_{t}) d P\left(\boldsymbol{\sigma}_{t}\right)\right]\\
&=\max \left[\int\limits_{\mathscr{C}_{1, t}}(\min \boldsymbol{b}^{\mathrm{T}} \boldsymbol{y}_{t}) d P\left(\boldsymbol{\sigma}_{t}\right)\right.
\\
&\phantom{=\;\;}
\left.+\sum_{s=1}^{M-1} \int\limits_{\mathscr{C}_{s+1, t}\backslash \mathscr{C}_{s, t}}(\min \boldsymbol{b}^{\mathrm{T}} \boldsymbol{y}_{t}) d P\left(\boldsymbol{\sigma}_{t}\right)\right]
\end{aligned}
\end{equation}
\end{small}

Similarly, the constraints (58) can be transformed as:
\begin{equation}
\begin{aligned}
&\left\{\begin{array}{l}
\int\limits_{\mathscr{C}_{1, t}}\boldsymbol{\sigma}_{t} d P\left(\boldsymbol{\sigma}_{t}\right)+\sum\limits_{s=1}^{M_{0}-1} \int\limits_{\mathscr{C}_{s+1, t} \backslash \mathscr{C}_{s, t}} \boldsymbol{\sigma}_{t} d P\left(\boldsymbol{\sigma}_{t}\right)=\mathbf{0}: \boldsymbol{\eta}_{t} \\
\int\limits_{\mathscr{C}_{1, t}} d P\left(\boldsymbol{\sigma}_{t}\right)+\sum\limits_{s=1}^{m-1} \int\limits_{\mathscr{C}_{s+1,t}, \mathscr{C}_{s, t}} d P\left(\boldsymbol{\sigma}_{t}\right)=P_{m, t}: \gamma_{mt}
\end{array}\right.
\\
&\left(\boldsymbol{\eta}_{t} \in \mathbb{R}^{N_{b}} \times \mathbb{R}^{N_{b}}, \forall m \in \mathcal{M}\right)
\end{aligned}
\end{equation}
where $\boldsymbol{\eta}_{t}$ and $\gamma_{mt}$ represent the dual vector and variable corresponding constraints, and $\boldsymbol{\eta}_{t}\ge\boldsymbol{0}$, $\gamma_{mt}\ge0$.

Based on the dual theory, the maximization problem (60) and (61) can be written as an equivalent minimization problem as follows:
\begin{equation}
\begin{aligned}
&\min \sum_{m=0}^{M_{0}} \gamma_{m t} P_{m, t}  \\
&\text { s.t. }\left(\boldsymbol{\sigma}_{t}\right)^{\mathrm{T}} \boldsymbol{\eta}_{t}+\sum_{s=1}^{M_{0}} \gamma_{s t} \geq \min \boldsymbol{b}^{\mathrm{T}} \boldsymbol{y}_{t}, \forall \boldsymbol{\sigma}_{t} \in \mathscr{C}_{1, t}\\
&\left(\boldsymbol{\sigma}_{t}\right)^{\mathrm{T}} \boldsymbol{\eta}_{t}+\sum_{s=m-1}^{M_{0}} \gamma_{s t} \geq \min \boldsymbol{b}^{\mathrm{T}} \boldsymbol{y}_{t}, \forall \boldsymbol{\sigma}_{t} \in \mathscr{C}_{s+1, t} \backslash \mathscr{C}_{s, t}, \\ &\ \ \ \ \ \ \ \ \ \ \ \ \ \ \ \ \ \  \ \ \ \ \ \ \ \ \ \ \ \ \ \ \ \ \ \ \ \ \ \ \ \ \ \ \ \ \ \ \ \ \forall m \in \mathcal{M} \backslash[1]
\end{aligned}
\end{equation}
Then the constraints in (62) can be transformed as
\begin{equation}
\sum_{s=m-1}^{M_{0}} \gamma_{s t} \geq \max_{\boldsymbol{\sigma}_{t} \in \mathscr{C}_{m, t}}[\min \boldsymbol{b}^{\mathrm{T}} \boldsymbol{y}_{t}-\left(\boldsymbol{\sigma}_{t}\right)^{\mathrm{T}} \boldsymbol{\eta}_{t}], \forall m \in \mathcal{M} \backslash[1]
\end{equation}

Combining with (55) and (63), the second stage of the bi-level programming problem can be reformulated as:
\begin{equation}
\begin{aligned}
&\max_{\boldsymbol{\sigma}_{t} \in \mathscr{C}_{m, t}}[\min \boldsymbol{b}^{\mathrm{T}} \boldsymbol{y}_{t}-\left(\boldsymbol{\sigma}_{t}\right)^{\mathrm{T}} \boldsymbol{\eta}_{t}]
\\
&\text { s.t. }\boldsymbol{B}_{t} \boldsymbol{x}+\boldsymbol{C}_{t} \boldsymbol{y}_{t} \leq \boldsymbol{d}_{t}+\boldsymbol{D}_{t} \boldsymbol{\sigma}_{t}:\ \boldsymbol{\nu}_{t}
\end{aligned}
\end{equation}
where $\forall m \in \mathcal{M} \backslash[1]$, and $\boldsymbol{\nu}_{t}$ is the dual vector of the second stage constraints, $\boldsymbol{\nu}_{t}\ge\boldsymbol{0}$.

According to the duality theory, (64) can be converted as a single-level optimization model as follows:
\begin{equation}
\begin{aligned}
&\max _{\boldsymbol{\sigma}_{t}, \boldsymbol{\nu}_{t}}\left[-\left(\boldsymbol{\sigma}_{t}\right)^{\mathrm{T}} \boldsymbol{\eta}_{t}+\boldsymbol{\nu}_{t}^{\mathrm{T}}\left(\boldsymbol{B}_{t} \boldsymbol{x}-\boldsymbol{d}_{t}-\boldsymbol{D}_{t} \boldsymbol{\sigma}_{t}\right)\right] \\
&\text { s.t. } \boldsymbol{b}=\left(\boldsymbol{C}_{t}\right)^{\mathrm{T}} \boldsymbol{\nu}_{t}
\end{aligned}
\end{equation}
\qed

Assuming two uncertain vectors $\boldsymbol{\sigma}_{1, t}$ and $\boldsymbol{\sigma}_{2, t}$ being defined as $\boldsymbol{0}\le\boldsymbol{\sigma}_{1, t}$,$\boldsymbol{\sigma}_{2, t}\le\boldsymbol{1}$, and $\boldsymbol{\sigma}_{1, t}, \boldsymbol{\sigma}_{2, t}\in\mathbb{R}^{N_{b}} \times \mathbb{R}^{N_{b}}$, the constraints of confidence set $\mathscr{C}_{m, t}$ can be converted as \cite{R. Li_A distributionally}

\begin{equation}
\left\{\begin{array}{l}
\boldsymbol{\sigma}_{t}=\boldsymbol{\sigma}_{1, t}-\boldsymbol{\sigma}_{2, t} \\
\mathbf{1}^{\prime} \cdot\left(\boldsymbol{\sigma}_{1, t}+\boldsymbol{\sigma}_{2, t}\right) \leq \Gamma_{m, t}
\end{array}\right.
\end{equation}

Then the bilinear term in the constraint of Proposition 3 can be written as

\begin{equation}
\begin{aligned}
&\max _{ \boldsymbol{\sigma}_{t} \in \mathscr{C}_{m, t}}\left(-\boldsymbol{\eta}_{t}^{\mathrm{T}}-\boldsymbol{v}_{t}^{\mathrm{T}} \boldsymbol{D}_{t}\right) \boldsymbol{\sigma}_{t} \\
&\text { s.t. } \boldsymbol{\sigma}_{t}=\boldsymbol{\sigma}_{1, t}-\boldsymbol{\sigma}_{2, t} \\
&\qquad\mathbf{0} \leq \boldsymbol{\sigma}_{1, t} \leq \mathbf{1}: \boldsymbol{\tau}_{1, t} \\
&\qquad\mathbf{0} \leq \boldsymbol{\sigma}_{2, t} \leq \mathbf{1}: \boldsymbol{\tau}_{2, t} \\
&\qquad\mathbf{1}^{\prime} \cdot\left(\boldsymbol{\sigma}_{1, t}+\boldsymbol{\sigma}_{2, t}\right) \leq \Gamma_{m, t}: \rho_{t}
\end{aligned}
\end{equation}

\noindent where $\boldsymbol{\tau}_{1, t},\boldsymbol{\tau}_{2, t}$ and $\rho_{t}$ represent the dual vectors and variable corresponding to the constraints, respectively.

Based on the dual theory, the maximization problem (67) can be written as an equivalent minimization problem as follows:
\begin{equation}
\begin{aligned}
&\min -\mathbf{1}^{\prime} \cdot\left(\boldsymbol{\tau}_{1, t}+\boldsymbol{\tau}_{2, t}\right)-\rho_{t} \Gamma_{m, t} \\
&\text { s.t. } \boldsymbol{\nu}_{t}^{\mathrm{T}} \boldsymbol{D}_{t}+\boldsymbol{\eta}_{t}^{\mathrm{T}}+\boldsymbol{\tau}_{1, t}+\mathbf{1} \cdot \rho_{t} \geq \mathbf{0} \\
&\qquad-\boldsymbol{\eta}_{t}^{\mathrm{T}}-\boldsymbol{\nu}_{t}^{\mathrm{T}} \boldsymbol{D}_{t}+\boldsymbol{\tau}_{2, t}+\mathbf{1} \cdot \rho_{t} \geq \mathbf{0} \\
&\qquad\boldsymbol{\tau}_{1, t}, \boldsymbol{\tau}_{2, t} \geq \mathbf{0}, \rho_{t} \geq 0
\end{aligned}
\end{equation}

Combined with (68), the single-level optimization model in the constraint of Proposition 3 can be converted as follows:

\begin{equation}
\begin{aligned}
&\max \boldsymbol{\nu}_{t}^{\mathrm{T}}\left(\boldsymbol{B}_{t} \boldsymbol{x}-\boldsymbol{d}_{t}\right)+\mathbf{1}^{\mathrm{T}} \cdot\left(\boldsymbol{\tau}_{1, t}+\boldsymbol{\tau}_{2, t}\right)+\rho_{t} \Gamma_{m, t} \\
&\text { s.t. } \boldsymbol{b}=\left(\boldsymbol{C}_{t}\right)^{\mathrm{T}} \boldsymbol{\nu}_{t} \\
&\qquad\boldsymbol{\nu}_{t}^{\mathrm{T}} \boldsymbol{D}_{t}+\boldsymbol{\eta}_{t}^{\mathrm{T}}+\boldsymbol{\tau}_{1, t}+\mathbf{1} \cdot \rho_{t} \geq \mathbf{0} \\
&\qquad-\boldsymbol{\eta}_{t}^{\mathrm{T}}-\boldsymbol{\nu}_{t}^{\mathrm{T}} \boldsymbol{D}_{t}+\boldsymbol{\tau}_{2, t}+\mathbf{1} \cdot \rho_{t} \geq \mathbf{0} \\
&\qquad\boldsymbol{\tau}_{1, t}, \boldsymbol{\tau}_{2, t} \geq \mathbf{0}, \rho_{t} \geq 0, \boldsymbol{\nu}_{t} \geq \mathbf{0}
\end{aligned}
\end{equation}

The complementary slackness conditions of the optimal problem (68) can be written as
\begin{equation}
\left\{\begin{array}{l}
\left(\boldsymbol{\nu}_{t}^{\mathrm{T}} \boldsymbol{D}_{t}+\boldsymbol{\eta}_{t}^{\mathrm{T}}+\boldsymbol{\tau}_{1, t}+\mathbf{1} \cdot \rho_{t}\right) \boldsymbol{\sigma}_{1, t}=\mathbf{0} \\
\left(-\boldsymbol{\eta}_{t}^{\mathrm{T}}-\boldsymbol{\nu}_{t}^{\mathrm{T}} \boldsymbol{D}_{t}+\boldsymbol{\tau}_{2, t}+\mathbf{1} \cdot \rho_{t}\right) \boldsymbol{\sigma}_{2, t}=\mathbf{0} \\
\left(\boldsymbol{\sigma}_{1, t}-\mathbf{1}\right)^{\mathrm{T}} \boldsymbol{\tau}_{1, t}=\mathbf{0} \\
\left(\boldsymbol{\sigma}_{2, t}-\mathbf{1}\right)^{\mathrm{T}}\boldsymbol{\tau}_{2, t}=\mathbf{0}
\end{array}\right.
\end{equation}

The nonlinear constraints in (70) can be converted by the big-M method as follows:

\begin{equation}
\left\{\begin{array}{l}
\mathbf{0} \leq \boldsymbol{\tau}_{1, t} \leq M \boldsymbol{\varepsilon}_{1, t} \\
\mathbf{0} \leq \boldsymbol{\tau}_{2, t} \leq M \boldsymbol{\varepsilon}_{2, t} \\
\mathbf{0} \leq \boldsymbol{\nu}_{t}^{\mathrm{T}} \boldsymbol{D}_{t}+\boldsymbol{\eta}_{t}^{\mathrm{T}}+\boldsymbol{\tau}_{1, t}+\mathbf{1} \cdot \rho_{t} \leq M\boldsymbol{\iota}_{1, t} \\
\mathbf{0} \leq-\boldsymbol{\eta}_{t}^{\mathrm{T}}-\boldsymbol{\nu}_{t}^{\mathrm{T}} \boldsymbol{D}_{t}+\boldsymbol{\tau}_{2, t}+\mathbf{1} \cdot \rho_{t} \leq M \boldsymbol{\iota}_{2, t} \\
\boldsymbol{\varepsilon}_{1, t} \leq \boldsymbol{\sigma}_{1, t} \leq \mathbf{1}-\boldsymbol{\iota}_{1, t} \\
\boldsymbol{\varepsilon}_{2, t} \leq \boldsymbol{\sigma}_{2, t} \leq \mathbf{1}-\boldsymbol{\iota}_{2, t}
\end{array}\right.
\end{equation}

Then the subproblem of the optimal problem (55) can be expressed as the following MILP problem:
\begin{equation}
\begin{aligned}
&\operatorname{SP}:  \max \boldsymbol{\nu}_{t}^{\mathrm{T}}\left(\boldsymbol{B}_{t} \boldsymbol{x}-\boldsymbol{d}_{t}\right)+\mathbf{1}^{\mathrm{T}} \cdot\left(\boldsymbol{\tau}_{1, t}+\boldsymbol{\tau}_{2, t}\right)+\rho_{t} \Gamma_{m, t} \\
&\text { s.t. } \boldsymbol{b}=\left(\boldsymbol{C}_{t}\right)^{\mathrm{T}} \boldsymbol{\nu}_{t} \\
&\qquad\mathbf{0} \leq \boldsymbol{\tau}_{1, t} \leq M \boldsymbol{\varepsilon}_{1, t} \\
&\qquad\mathbf{0} \leq \boldsymbol{\tau}_{2, t} \leq M \boldsymbol{\varepsilon}_{2, t} \\
&\qquad\mathbf{0} \leq \boldsymbol{\nu}_{t}^{\mathrm{T}} \boldsymbol{D}_{t}+\boldsymbol{\eta}_{t}^{\mathrm{T}}+\boldsymbol{\tau}_{1, t}+\mathbf{1} \cdot \rho_{t} \leq M \boldsymbol{\iota}_{1, t} \\
&\qquad\mathbf{0} \leq -\boldsymbol{\eta}_{t}^{\mathrm{T}}-\boldsymbol{\nu}_{t}^{\mathrm{T}} \boldsymbol{D}_{t}+\boldsymbol{\tau}_{2, t}+\mathbf{1} \cdot \rho_{t} \leq M \boldsymbol{\iota}_{2, t} \\
&\qquad \boldsymbol{\varepsilon}_{1, t} \leq \boldsymbol{\sigma}_{1, t} \leq \mathbf{1}-\boldsymbol{\iota}_{1, t}\\
&\qquad \boldsymbol{\varepsilon}_{2, t} \leq \boldsymbol{\sigma}_{2, t} \leq \mathbf{1}-\boldsymbol{\iota}_{2, t} \\
&\qquad\mathbf{1}^{\prime} \cdot\left(\boldsymbol{\sigma}_{1, t}+\boldsymbol{\sigma}_{2, t}\right) \leq \Gamma_{m, t}
\end{aligned}
\end{equation}
where $\boldsymbol{\varepsilon}_{1, t}$, $\boldsymbol{\varepsilon}_{2, t}$, $\boldsymbol{\iota}_{1, t}$ and $\boldsymbol{\iota}_{2, t}$ denote binary variables.

Next, we implement the Benders decomposition algorithm to solve the DRO model (55) \cite{M. Zugno_A Robust}. According to Ref. \cite{A.M.Geoffrion_Generalized}, the Benders cut is expressed as

\begin{equation}
\sum_{s=m-1}^{M_{0}} \gamma_{s t} \geq-\left(\boldsymbol{\sigma}_{m t}^{*}\right)^{\mathrm{T}} \boldsymbol{\eta}_{t}+\left(\boldsymbol{\nu}_{m t}^{*}\right)^{\mathrm{T}}\left(\boldsymbol{B}_{t} \boldsymbol{x}-\boldsymbol{d}_{t}-\boldsymbol{D}_{t} \boldsymbol{\sigma}_{m t}^{*}\right)
\end{equation}

\noindent where $\boldsymbol{\sigma}_{t}^{*}$ and $\boldsymbol{\nu}_{t}^{*}$ are the optimal solution of the $\rm SP$. Then the $\rm MP$ can be written as
\begin{equation}
\mathrm{MP}: \min \left(\boldsymbol{c}^{\mathrm{T}} \boldsymbol{x}+\sum_{t \in \mathcal{T}} \sum_{m=0}^{M_{0}} \gamma_{m t} P_{m, t}\right) \notag~~~~~~~~~~~~~~~~~~~~~~~~
\end{equation}
\begin{equation}
\begin{aligned}
&\text { s.t. } \boldsymbol{A} \boldsymbol{x} \leq \boldsymbol{h} \\
&\sum_{s=m-1}^{M_{0}} \gamma_{s t} \geq-\left(\boldsymbol{\sigma}_{m t}^{* k}\right)^{\mathrm{T}} \boldsymbol{\eta}_{t}+\left(\boldsymbol{\nu}_{m t}^{* k}\right)^{\mathrm{T}}\left(\boldsymbol{B}_{t} \boldsymbol{x}-\boldsymbol{d}_{t}-\boldsymbol{D}_{t} \boldsymbol{\sigma}_{m t}^{* k}\right)
\end{aligned}
\end{equation}

\noindent where $\boldsymbol{\sigma}_{m t}^{* k}$ and $\boldsymbol{\nu}_{m t}^{* k}$ are the optimal solution of the $\rm SP$ at iteration $k$.

On this basis, the solving procedure of the DRO model (55) with a master-subproblem framework relying on the Benders decomposition algorithm is presented in Algorithm 1, and the parameter setting of the ambiguity set can affect the precision of optimal solution.

\begin{algorithm}[htbp]
\caption{Benders decomposition algorithm on DRO model.}\label{alg:alg1}
\begin{algorithmic}
\STATE 
\STATE {\textbf{Result}: Obtain the optimal day-ahead power scheduling strategy}
\STATE 1. Initialization. Let the lower bound $LB=-\infty$, upper bound $UB=+\infty$ and iteration number $k=0$. Choose a convergence tolerance level $\xi>0$. Fix a feasible solution $(\boldsymbol{x}^{* k},\gamma_{m t}^{* k},\boldsymbol{\eta}_{t}^{* k})$ of $\rm MP$.
\STATE 2. Solve the $\rm SP$ related to $(\boldsymbol{x}^{* k},\gamma_{m t}^{* k},\boldsymbol{\eta}_{t}^{* k})$, to get the objective value $ \sum\limits_{s=m-1}^{M_{0}} \gamma_{s t}^{*k}$ and optimal solution $(\boldsymbol{\sigma}_{m t}^{* k},\boldsymbol{\nu}_{m t}^{* k})$.
\STATE 3. Obtain the objective value $OB=\boldsymbol{c}^{\mathrm{T}} \boldsymbol{x}^{*k}+\sum\limits_{t \in \mathcal{T}} \sum\limits_{m=0}^{M_{0}}\sum\limits_{s=m-1}^{M_{0}} \gamma_{s t}^{*k}P_{m,t}$ of $\rm MP$ with respect to $ \boldsymbol{x}^{*k}$ and $\sum\limits_{s=m-1}^{M_{0}} \gamma_{s t}^{*k}$, and let $UB$$=$$\text{min}\{OB,UB\}$.
\STATE 4. Update the Benders cut related to $(\boldsymbol{\sigma}_{m t}^{* k},\boldsymbol{\nu}_{m t}^{* k})$ and solve $\rm MP$. Then update the optimal solution and objective value to $(\boldsymbol{x}^{* (k+1)},\gamma_{m t}^{* (k+1)},\boldsymbol{\eta}_{t}^{* (k+1)})$ and $z^{*(k+1)}$, respectively. Let $LB=z^{*(k+1)}$.
\STATE 5. If $(UB-LB)/UB\le\xi$, terminate the procedure and return $\boldsymbol{x}^{* k}$ as the optimal solution. Otherwise, let $k=k+1$ and go to Step 2.
\end{algorithmic}
\label{alg1}
\end{algorithm}

\section{Case Study}
In this section, numerical analysis is performed for the 24-h scheduling in an 8-MG power system coupled with a benchmark urban TN, as depicted in Fig. 4 and Fig. 5. The distribution network is modified from an IEEE 33-bus system with PVs. 8-MG is located in the power system, and is connected with one bus, respectively. The detailed data can be referred to \cite{H. Gao_Robust coordinated}. The arrows of each link represent the direction of vehicles permitted to drive. This urban TN is widely utilized in the research related to power and transportation coupled networks (e.g., \cite{W. Gan_Coordinated}, \cite{W. Wei_Robust}, \cite{Y. Liu_Multi-Agent}). Relevant parameters of traffic links are listed in Table \uppercase\expandafter{\romannumeral2}. It is noted that the proposed approach is extensible, which can be applied to the urban TN with bidirectional-link and multiple O-D pairs. Owing to the spatial limitation of this paper, only three O-D pairs in this transportation system are considered, and the parameters are listed in Table \uppercase\expandafter{\romannumeral3}, in which the traffic demand is the average value from the historical data \cite{W. Wei_Robust} and the basic value (p.u.) is 100 vehicles per hour. The simulations are implemented on a laptop with an Intel Core i9-10885H CPU 2.40 GHz using MATLAB with YALMIP and CPLEX 12.9.0 solver.

\begin{figure}[htbp]
 \centering
\includegraphics[width=3in]{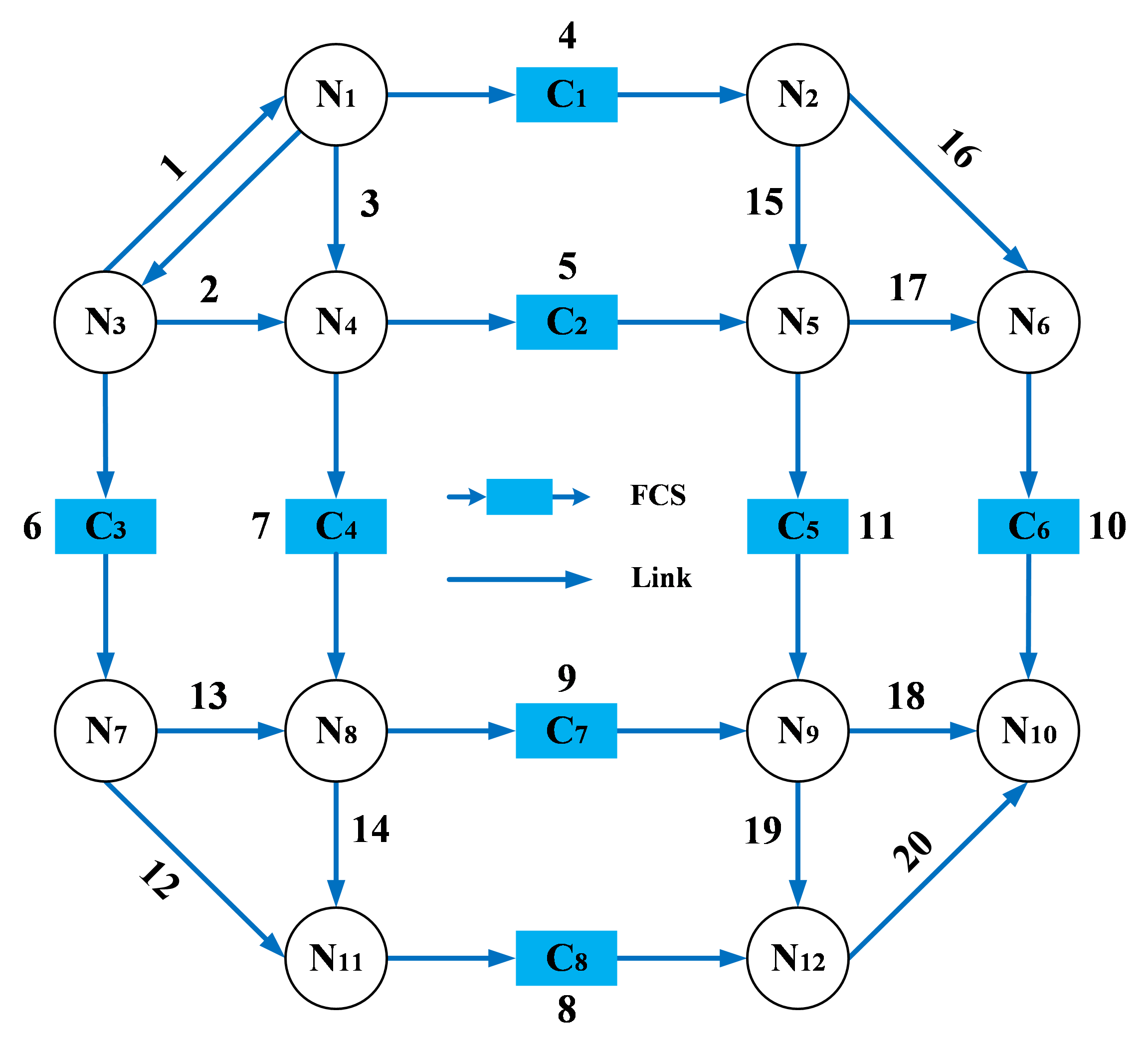}
\caption{Topology of TN with FCS.}
\label{figl}
 \end{figure}

 \begin{figure}[htbp]
 \centering
\includegraphics[width=3in]{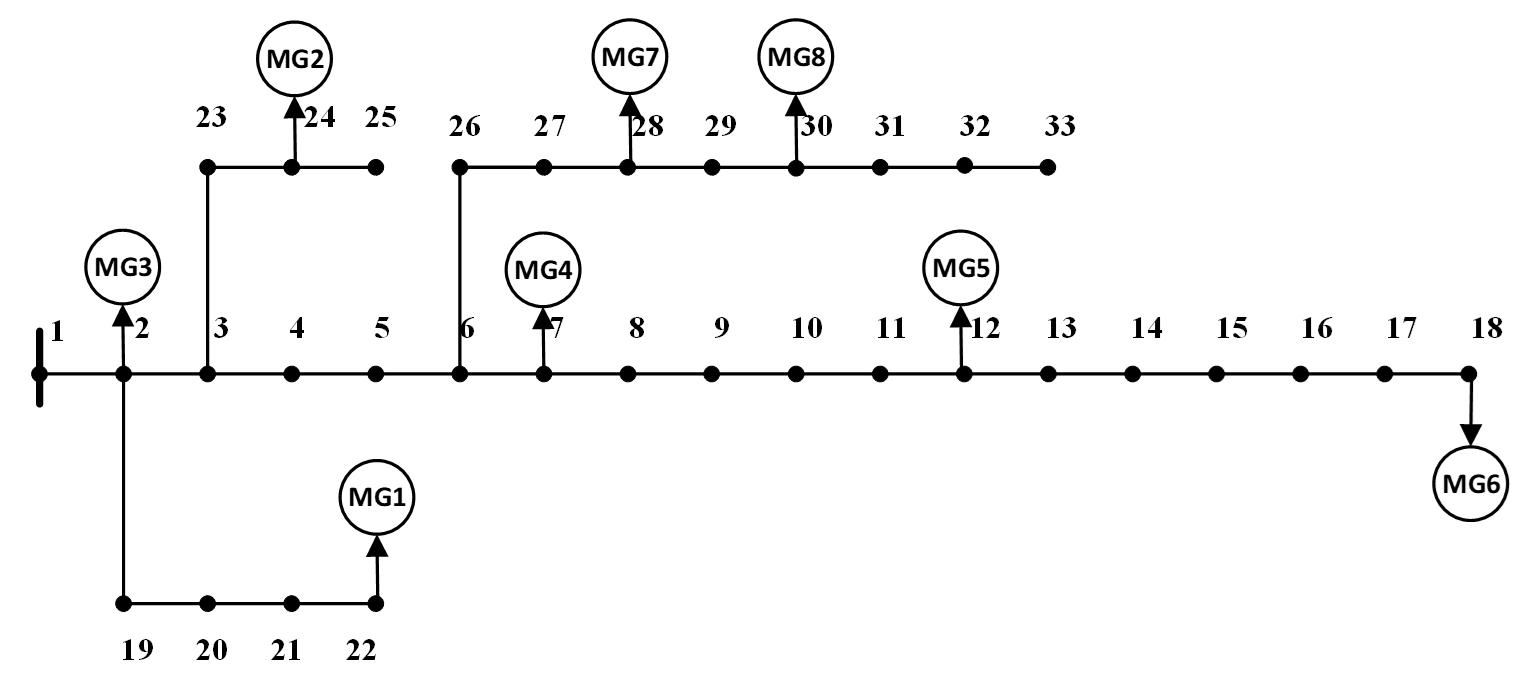}
\caption{Topology of distribution network.}
\label{figl}
\end{figure}
 
\begin{table}[htbp]
\caption{Parameters for the 20-links transportation system\label{tab:table1}}
\centering
\begin{tabular}{cccccc}
\hline
link & $C_{a}(\rm p.u.)$  & $t_{l}^{0}(min)$ & link & $C_{a}(\rm p.u.)$ & $t_{l}^{0}(min)$ \\
\hline
1  & 18 & 6 & 11 & 13.8 & 12 \\
2  & 8.5 & 6 & 12 & 17.5 & 6 \\
3  & 9.8 & 5 & 13 & 8.9 & 5 \\
4  & 20 & 10 & 14 & 9.76 & 5 \\
5  & 13.5 & 12 & 15 & 7.9 & 5 \\
6  & 19 & 10 & 16 & 17 & 6.5 \\
7  & 14 & 11 & 17 & 8.2 & 6.5 \\
8  & 20 & 9 & 18 & 9.15 & 5.9 \\
9  & 13.2 & 11 & 19 & 8.97 & 5.8 \\
10  & 20 & 10 & 20 & 18.2 & 6.1 \\
\hline
\end{tabular}
\end{table}

\begin{table}[htbp]
\caption{Parameters for O-D pairs\label{tab:table1}}
\centering
\begin{tabular}{cccc}
\hline
O-D pair & From node  & To node & \multicolumn{1}{m{2cm}}{Average traffic demand (p.u.)} \\
\hline
1-6  & N1 & N6 & 15  \\
3-11  & N3 & N11 & 12  \\
4-12  & N4 & N12 & 15  \\
\hline
\end{tabular}
\end{table}

In the power network, MG$1$-MG$8$ serves FCSs of C$1$-C$8$, respectively. We assume that the charging power of each EV is a constant $0.015\rm MWh$ and the monetary value of travel time $\omega$ is 10\$/h. The relevant parameters of MGs are listed in Table \uppercase\expandafter{\romannumeral4}, where two types of ES and DG devices are set according to the internal and external circulation flows of the transportation network. The ladder electricity price shown in Fig. 6 is used as the trading price between the main grid and MGs.

\begin{table}[htbp]
\caption{Operation parameters of MGs\label{tab:table1}}
\centering
\begin{tabular}{ccc}
\hline
\hline
Unit & Parameter  & Value \\
\hline
\multicolumn{1}{m{2cm}}{Exchange power with main grid} & $P_{\max }^{\rm G}$$(\rm MW)$ & 30 \\
\hline
\multirow{3}{*}{$\rm DG \_ \rm \uppercase\expandafter{\romannumeral1}$} & $P_{\max }^{\rm DG}(\rm MW)$ & 10 \\ 
& $P_{\min }^{\rm DG}(\rm MW)$ & 0.5 \\
& $a/b/c(\$ / \rm MW)$ & $0.1/106/0$\\
\hline
\multirow{3}{*}{$\rm DG \_ \rm \uppercase\expandafter{\romannumeral2}$} & $P_{\max }^{\rm DG}(\rm MW)$ & 20 \\ 
& $P_{\min }^{\rm DG}(\rm MW)$ & 2 \\
& $a/b/c(\$ / \rm MW)$ & $0.1/106/0$\\
\hline
\multirow{4}{*}{$\rm ES \_ \rm \uppercase\expandafter{\romannumeral1}$} & $\lambda^{\rm{E S}}(\$ / \rm{MW})$ & 60 \\ 
& $P_{\max }^{\rm{E S}}(\rm{MW})$ & 10 \\
& $S O C_{\max } \cdot E_{\rm{L}} (\rm{MWh})$ & $30$\\
& $S O C_{\min } \cdot E_{\rm{L}} (\rm{MWh})$ & $5$\\
\hline
\multirow{4}{*}{$\rm ES \_ \rm \uppercase\expandafter{\romannumeral2}$} & $\lambda^{\rm{E S}}(\$ / \rm{MW})$ & 60 \\ 
& $P_{\max }^{\rm{E S}}(\rm{MW})$ & 15 \\
& $S O C_{\max } \cdot E_{\rm{L}} (\rm{MWh})$ & $40$\\
& $S O C_{\min } \cdot E_{\rm{L}} (\rm{MWh})$ & $7$\\
\hline
Demand response & $\lambda_{\rm{D R}}(\$ / \rm{MW})$ & 50 \\
\hline
\hline
\end{tabular}
\end{table}

 \begin{figure}[htbp]
 \centering
\includegraphics[width=3in]{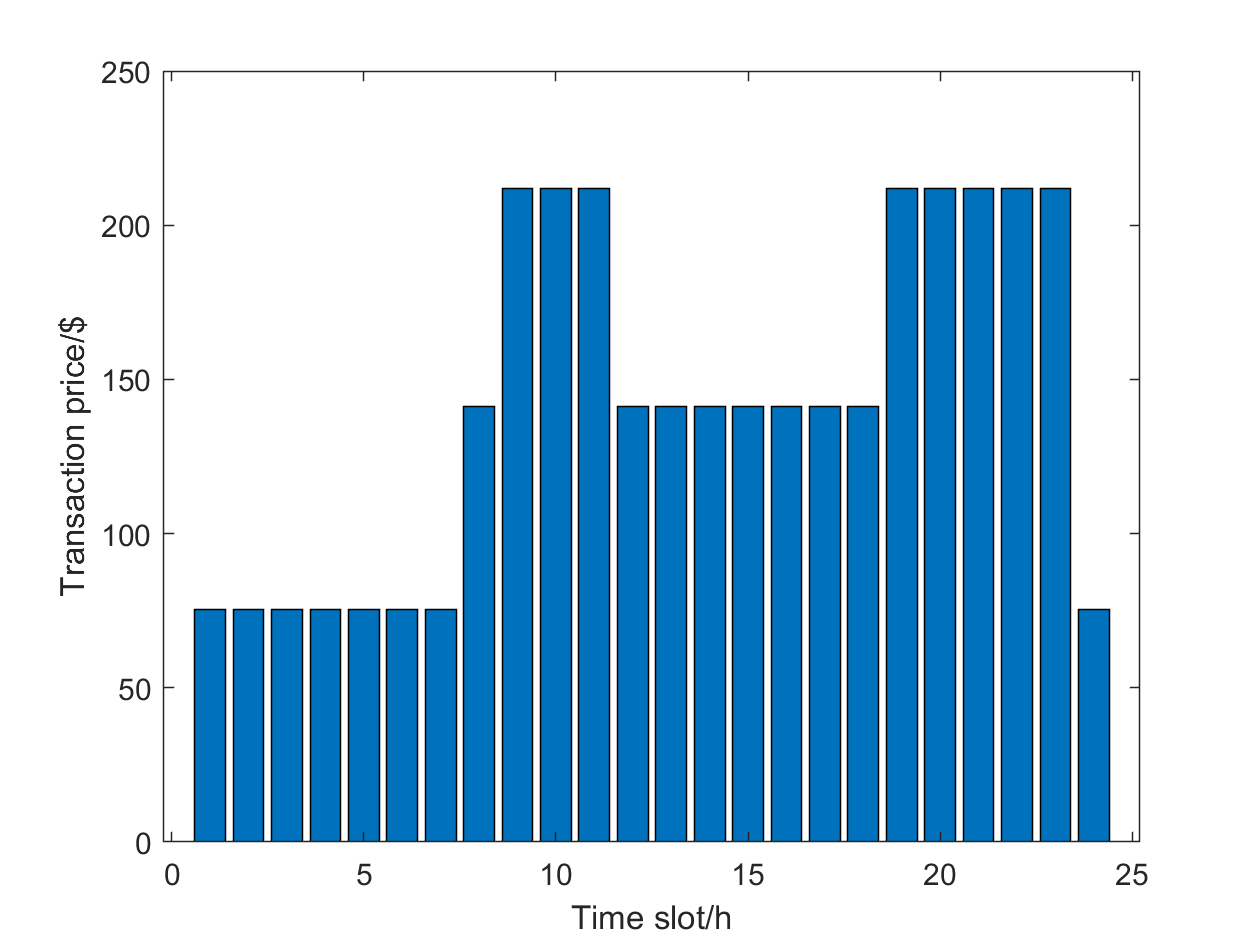}
\caption{The trading price between the main grid and MGs.}
\label{figl}
\end{figure}

In this paper, there are three kinds of cases being adopted, including the DRO model proposed by this paper, the RO model solved by C\&CG method \cite{B. Zeng_Solving}, and the deterministic model (DM) without uncertain variables, to research the influence on the conservativeness and economic performance of scheduling profile. In the DRO model case, the number of the confidence sets is set as 6, i.e., $M_{0}=5$. It should be noted that the model complexity and the computing time will increase with the rising of  $M_{0}$. While if a small value is assigned to $M_{0}$, the statistical features of uncertain parameters cannot be represented adequately. Hence, the value of $M_{0}$, the corresponding uncertainty budgets $\Gamma_{m,t}$ and probabilities $P_{m,t}$ are set by the distribution information of historical data. Meanwhile, the maximum allowable fluctuation deviation of the traffic demand and PV generation is set according to the previous historical prediction deviation, and the box sets are set as $\{ 0.9{q}_{t,av}^{od} \le \widetilde{q}_{t}^{od} \le 1.1{q}_{t,av}^{od}, \forall od \in \mathcal{R} \}$ and $\{ 0.85{p}_{j,t,av}^{\rm PV} \le \widetilde{p}_{j,t}^{PV} \le 1.15{p}_{j,t,av}^{\rm PV}, \forall j \in \mathcal{B} \}$, respectively. Then the charging demand of FCSs can be obtained by (45)-(48). In the RO case, the uncertainty set intervals are the same as the DRO model. In the DM case, the uncertainties of this coupled network are not considered. The average value in all time slots of PV generation is adopted. The UE traffic assignment is determined as a certain variable by solving (12) with the average traffic demand computed from historical data. Meanwhile, the DM optimization model is solved using a mixed integer linear programming method. The UE link flow pattern obtained by (12) is shown in Fig.7, in which the flow of $10^{th}$, $13^{th}$, $15^{th}$, $18^{th}$, and $20^{th}$ links is zero.
 
\begin{figure}[htbp]
\centering
\includegraphics[width=3in]{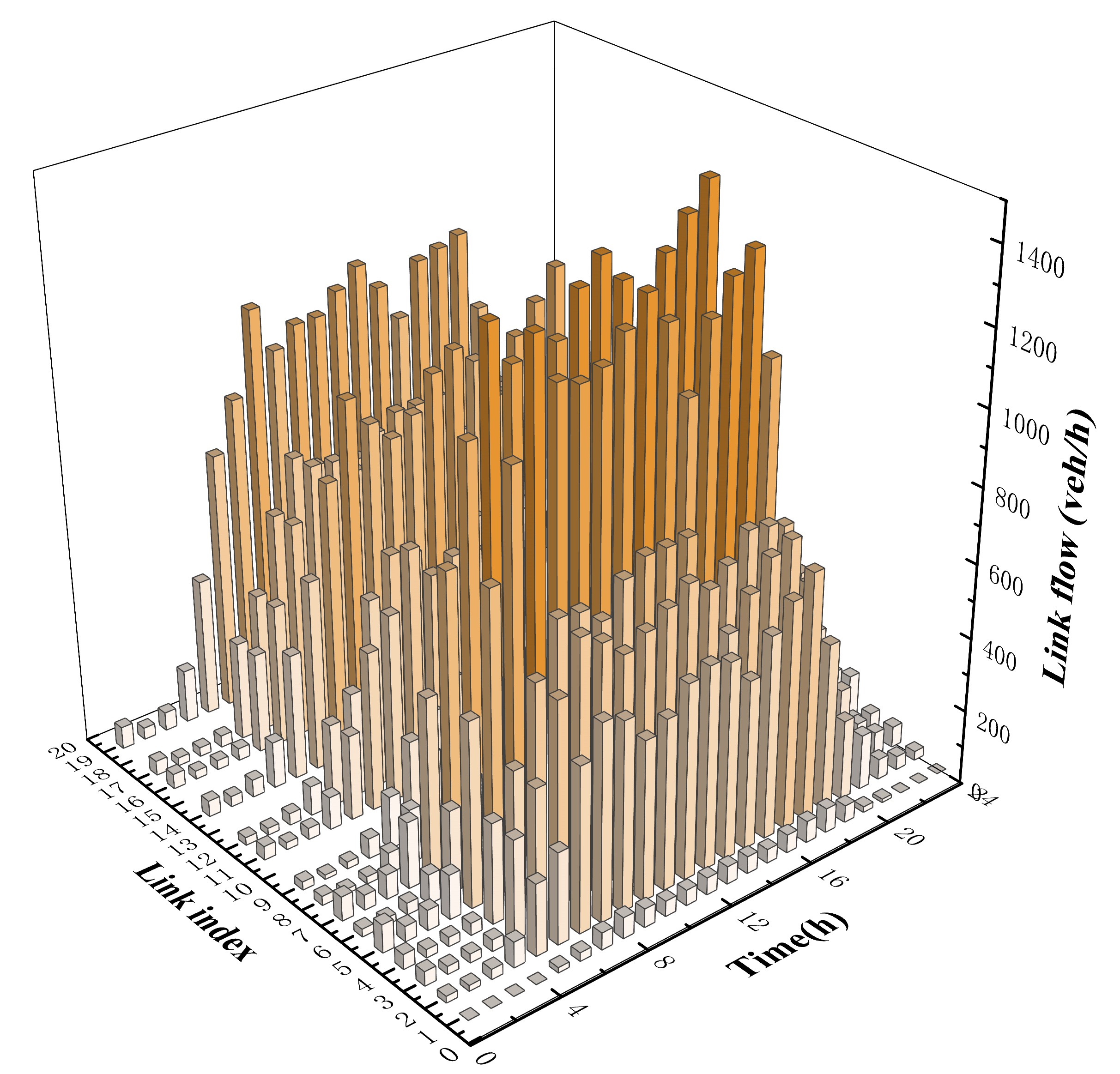}
\caption{The link flow with the average traffic demand in the total time slot.}
\label{figl}
\end{figure}
 
Taking MG$1$ and MG$4$ as examples, the optimization results are shown in Fig. 8-12. Note that the two types of ES and EG devices are employed by MG$1$ and MG$4$, respectively. The Fig. 8 and Fig. 10 reveal the energy scheduling optimization results of MG$1$ and MG$4$ by the DRO method, including the power output of DG, the exchange power of MG with main grid, the PV output, charging and discharging power of ES. It should be noted that when the MG purchases electricity from the main grid, the value of power is positive, otherwise, it is negative. Besides, the power value is negative when ES is charging, otherwise, it is positive. The actual dispatching of demand response loads and expected loads in all time slot are shown in Fig. 9 and Fig. 11.
  
\begin{figure}[htbp]
\centering
\includegraphics[width=3in]{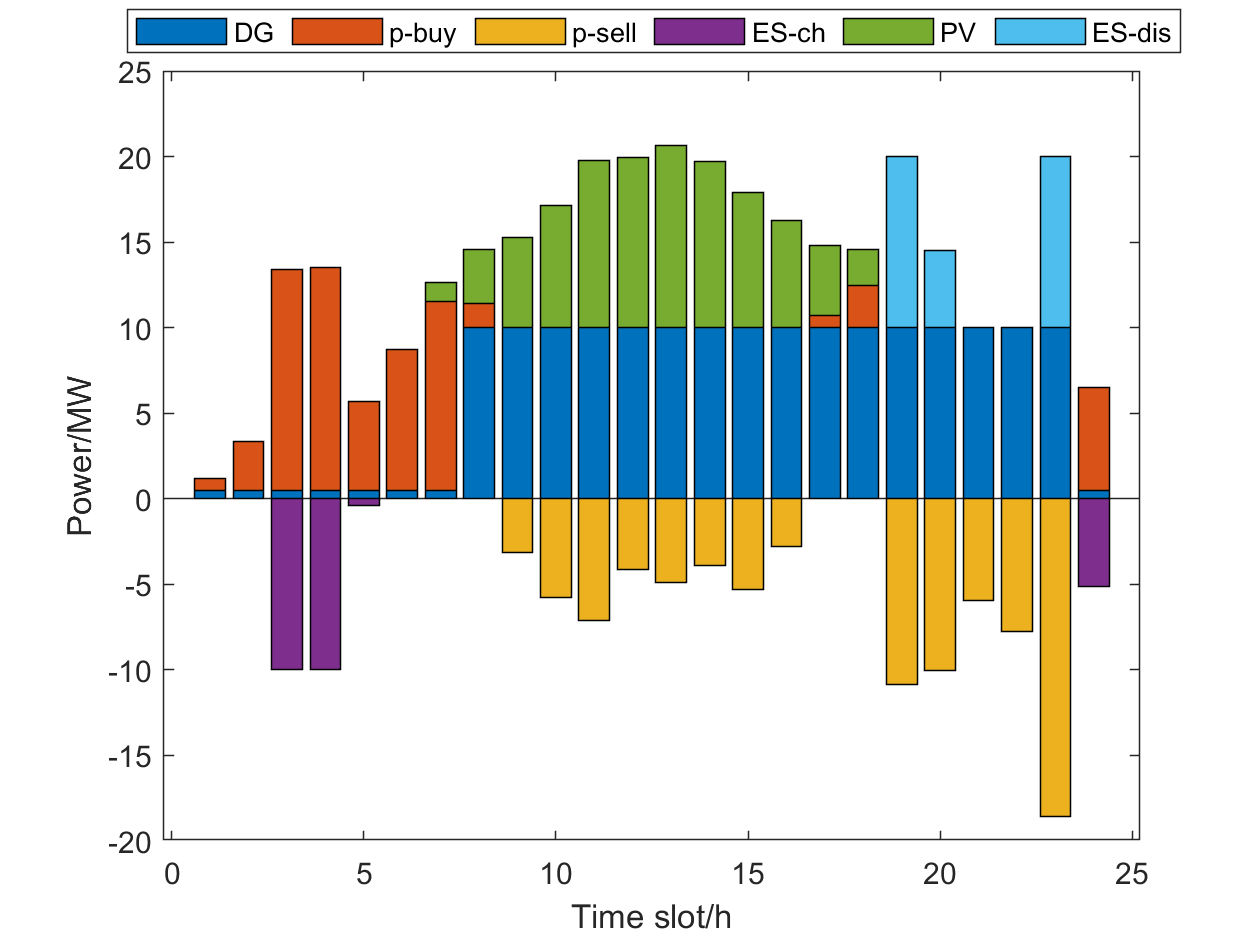}
\caption{Power scheduling profile by DRO for MG$1$.}
\label{figl}
\end{figure}

\begin{figure}[htbp]
\centering
\includegraphics[width=3in]{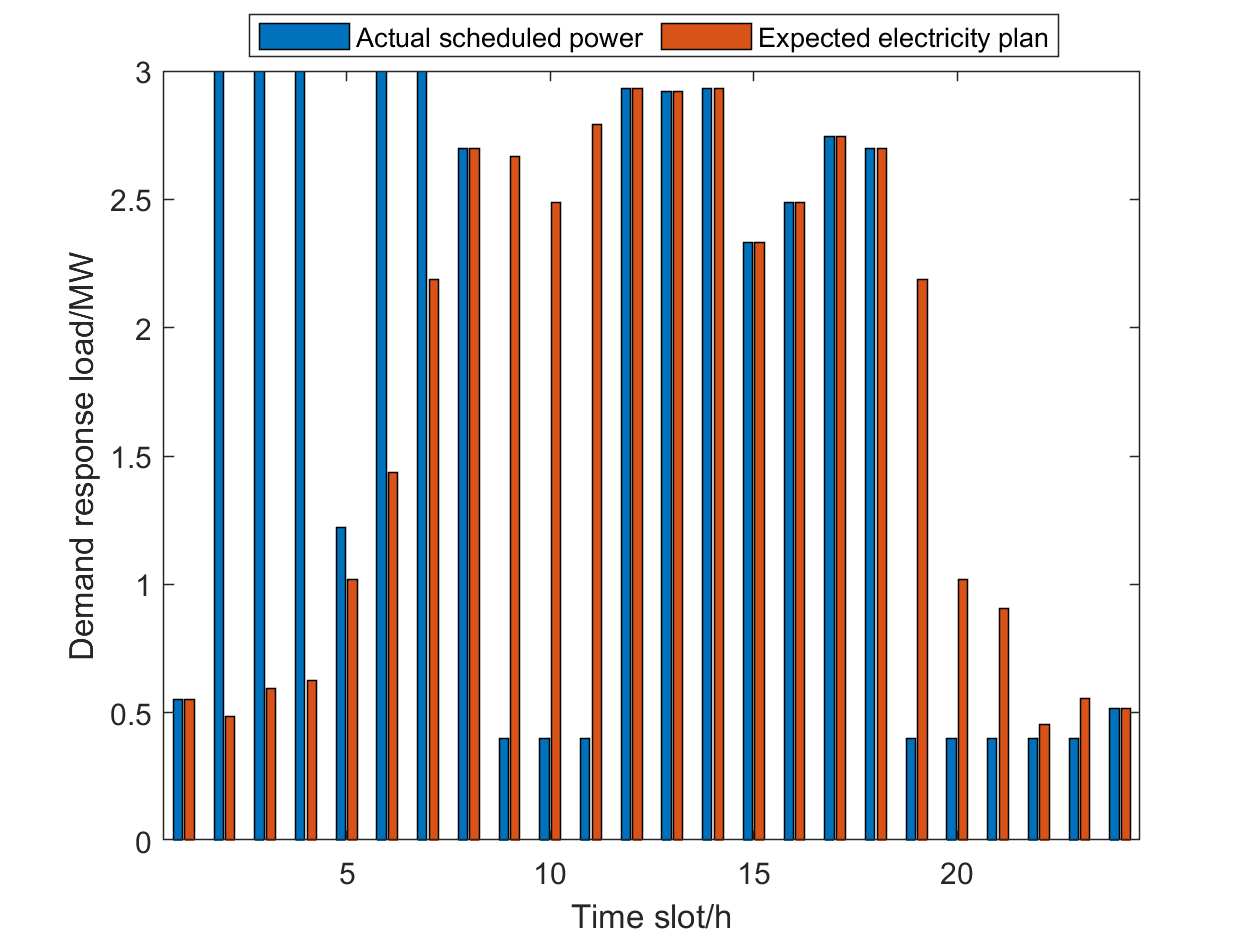}
\caption{Actual and expected electricity plan of demand response load by DRO for MG$1$.}
\label{figl}
\end{figure}

\begin{figure}[htbp]
\centering
\includegraphics[width=3in]{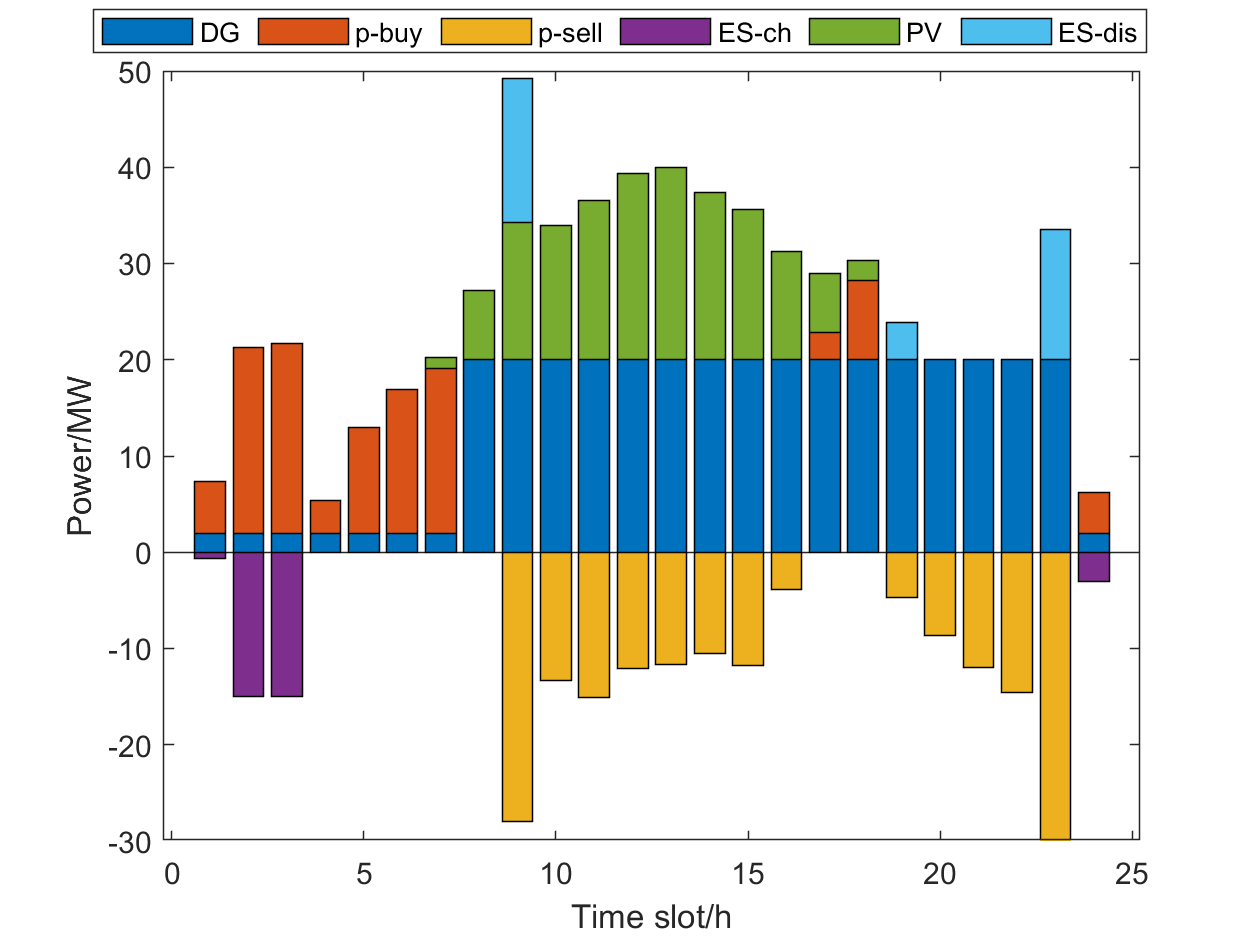}
\caption{Power scheduling profile by DRO for MG$4$.}
\label{figl}
\end{figure}

\begin{figure}[htbp]
\centering
\includegraphics[width=3in]{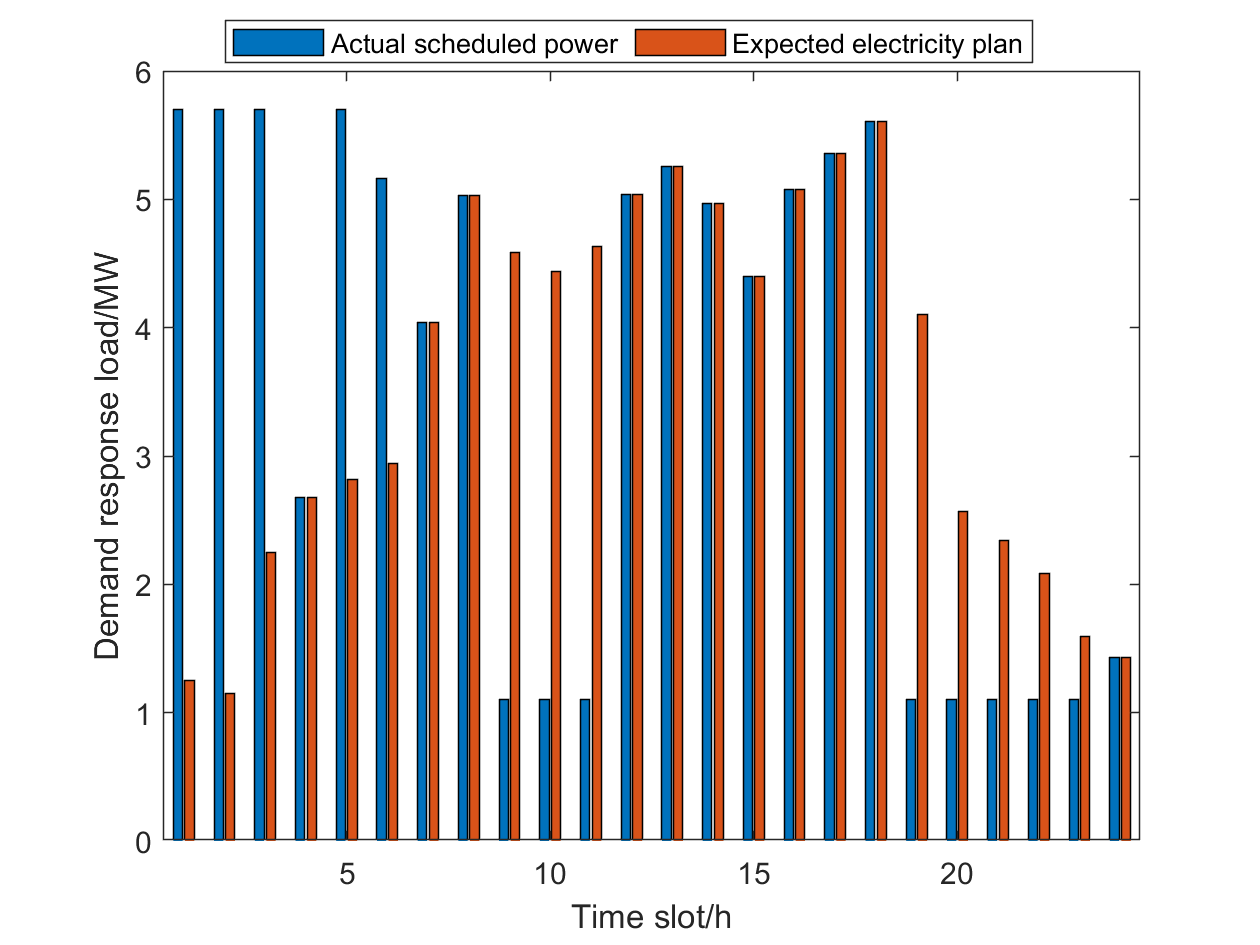}
\caption{Actual and expected electricity plan of demand response load by DRO for MG$4$.}
\label{figl}
\end{figure}

In Fig. 8 and Fig. 10, it can be seen that the PV output power is 0 in 1h\textasciitilde6h and 19h\textasciitilde24h, and the load in the MG is completely supplied by DG, ES and the power purchase from main grid. In this time period, the day-ahead transaction price is lower than the unit power generation cost of DG, and thus DG operates at the minimum output power point. During the rest of the time period, DG operates at the maximum power point to increase the power sold to the main grid, since the day-ahead transaction price is higher than the unit power generation cost of DG. Under the ladder electricity price mechanism, ES is charged during non-peak electricity consumption time period, and discharged at night or in the early morning. Then ES reserves the electricity during the valley electricity price period and sells it during the peak electricity price period. Therefore, the peak shaving and valley filling can be realized, which enhances the flexibility of the PN scheduling and makes the experimental results closer to the actual situation. In Fig. 9 and Fig. 11, the demand response load, which is similar to the traditional load, is mainly concentrated in the peak period of electricity price. Under the premise of satisfying the total power demand and the minimum power consumption in each time period, MG supplies excess power to demand response users during non-peak electricity consumption time period, which is stored to supply the power shortage during peak electricity consumption time period. This mechanism reduces the power purchase by MG from the main grid during the peak electricity price period and the cost of electricity consumption.

The situation of the power purchase and the power sold from/to the main grid of MG$1$ and MG$4$ under the three kinds of cases in one scheduling period are shown in Fig. 12 and Fig.13. In the DRO and RO methods, the uncertainty of traffic demand influences the UE state, so further causes the uncertainty of charging demand to MG and generates the reservation of charging power for EVs. Moreover, the worst case of uncertainties is considered in the RO method. As a result, the day-ahead power purchase from the main grid of MG applying DRO method is less than that of RO, and higher than DM on the general trend. In addition, the power outputs of PV generation by DRO method are more than the ones by RO, and thus the power sold to the main grid of MG using DRO method is greater than RO and lower than DM overall. In the DRO model proposed by this paper, the probability distribution for uncertain parameters can be captured based on the historical data. In contrast, the RO model adopts the boundary information of uncertainty box sets to obtain the optimal solution in the worst case increasing the conservativeness of strategies. Meanwhile, the DRO model can avoid the dilemma of selling more surplus electricity back to the main grid.

\begin{figure}[htbp]
\centering
\includegraphics[width=3in]{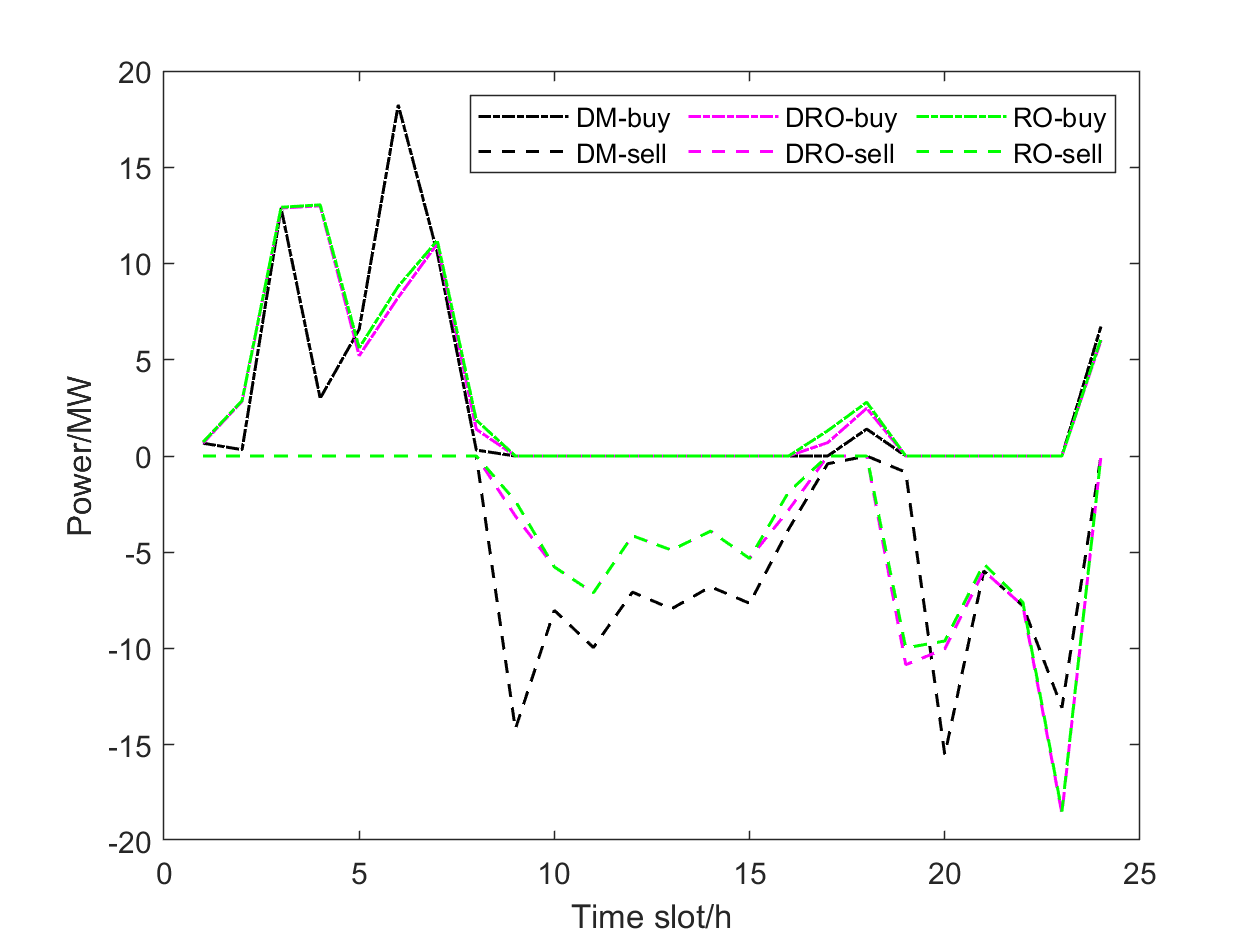}
\caption{The exchange power of MG$1$ with main grid by DRO, RO and DM.}
\label{figl}
\end{figure}

\begin{figure}[htbp]
\centering
\includegraphics[width=3in]{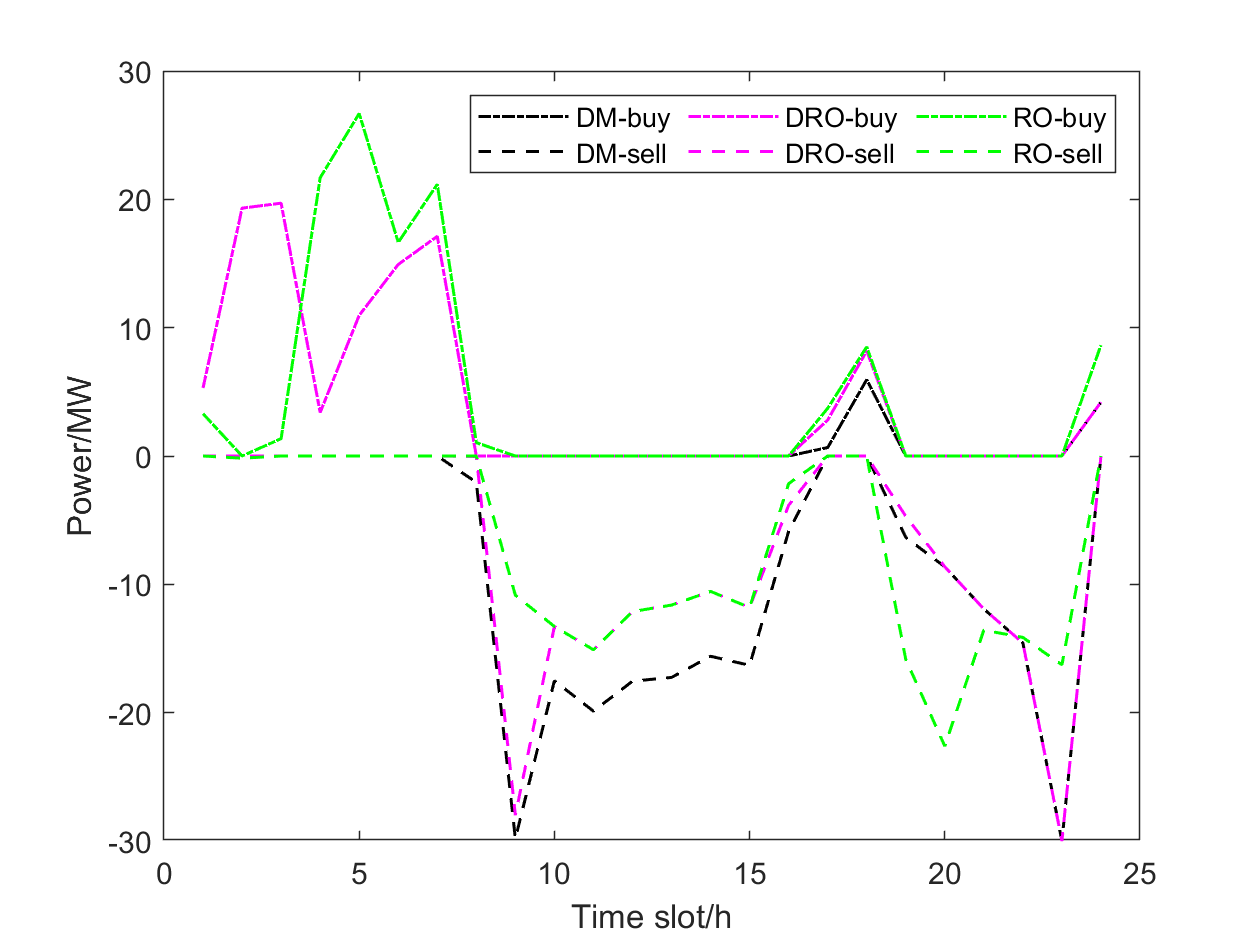}
\caption{The exchange power of MG$4$ with main grid by DRO, RO and DM.}
\label{figl}
\end{figure}

Table \uppercase\expandafter{\romannumeral5} provides the total cost , total power purchase and power sold from/to the main grid of the 8-MG system in three cases. As is shown in this table, the operating costs of the DRO and RO model are more than that of DM. Nonetheless, it is not mean that the scheme of the DM optimization method is better than the robust method. The reason is that this scheme corresponds to the generation and consumption plan submitted by the MG in the day-ahead market, and the imbalance between the planned generation and consumption caused by the uncertainties in the real-time market. The power purchase price in the real-time market is generally higher than the day-ahead market, which results in the increase of transaction cost for the MG. The robust approach considers the worse situation of the higher-level charging load and the lower-level PV generation, which leads to the more conservative strategies and the higher operating cost for the day-ahead schedule. Meanwhile, the uncertain factors can change the flow patterns and the route choices of vehicles, and the results of the game between EVs and MG operators are changed, which is neglected in DM. In addition, as is shown in Fig.12 and Fig. 13, the increase of operating costs is mainly generated by the increase of the power purchase and the decrease of power sold from$/$to the main grid. According to this point, the scheduling profile obtained by the robust optimization method has stronger robustness to resist the risk of real-time market loads and price fluctuation. Similarly, it can also be observed that the DRO method can obtain an excellent scheduling strategy with more economic performance compared with RO, which indicates that the DRO model proposed by this paper can reduce the conservativeness of strategies and total cost.

\begin{table}[htbp]
\caption{Simulation results in different cases\label{tab:table1}}
\centering
\begin{tabular}{cccc}
\hline
\hline
Model & RO  & DRO & DM \\
\hline
$P^{buy}$ $(\times10^{3}\rm MW)$  & 0.7199 & 0.6813 & 0.6488  \\
$P^{sell}$ $ (\times10^{3}\rm MW)$  & 1.0267 & 1.0657 & 1.2893 \\
Total cost $(\times10^{4}\$)$  & 11.1954 & 9.0474 & 5.9251 \\
\hline
\end{tabular}
\end{table}

Furthermore, about the performance of the above three cases, the total computing times of the DRO, RO, and DM are 187s, 96s, and 24s, respectively. As expected, the DM has a minimal calculation time. It should be noted that the DRO requires a longer computing time than RO, due to there being more auxiliary variables in the computing process of DRO. However, it can attain more practical scheduling and can adapt to multiple uncertainties.

Finally, we compare the optimization results of our method with the traditional operation case. In the traditional operation case, the individual selfishness and the game relationship between the MGs and the EV drivers are not considered. The optimization goal is modified to optimize the total benefit of the system in (75).
\begin{equation}
\begin{aligned}
&\min \ F_{\rm TN}(f_{p,t}^{od})+R_{j}\left(\boldsymbol{p}_{j, t}\right)\\
&\text{s.t.} (2)-(7), (9)-(11), (14), (15), (17)-(27), (29)-(37)
\end{aligned}
\end{equation}

In the traditional operation case, the solution method and parameter setting are the same as the DRO model of this paper. According to the simulation calculation, the total cost and the power sold from the main grid are $9.4760\times10^{4}\$$ and $0.7023\times10^{3}\rm MW$. It can be seen that the system operation cost of applying the traditional operation method is slightly lower than the method proposed by this paper, nevertheless, it does not take the intra-day game relationship and the individual selfishness of the coupled system in the optimization problem into account. It can lead to insufficient real-time power dispatching and the unmet EV charging requirements. The framework in this paper fully considers the individual selfishness without sacrificing the total cost, and the individual objectives can be achieved by optimizing the potential function.

\section{Conclusion}
The purpose of this study is to investigate day-ahead power scheduling for the distribution network while considering the coupling of PN and TN, in which the model construction of intra-day source-load uncertainties and the coupling expression of the two networks are the cruxes. The coupling characteristics of conflicting revenues and interaction accurately are characterized by the game theory, and then we prove that the game relation fulfills the characteristics of potential game. Furthermore, the DRO model considering the probability distribution of uncertain parameters is designed to derive the equilibrium of the potential game in the worst situation over ambiguity sets. It can reduce the conservativeness of the optimization model and ensure the safety and economy of power strategies effectively. Therefore, we propose the two-stage DRO model in a unified centralized optimization framework to carry out the day-ahead power scheduling and ensure the individual optimality. Simulation results confirm that the capability of DRO method based on potential game-theory in this paper is superior to the robust optimization and deterministic optimization approach. Meanwhile, the individual optimal strategies can be obtained by the approach of this paper without sacrificing the total cost. The limitation of this study is that there are some assumptions in the TN, which restrict the practical application of this method. As a result, the future extension of this investigation is to take the endogenous uncertainty for the travelers’ charging demand into account to enhance the application value in practice.

%
%
%
%
%

\end{document}